\documentclass[fleqn,usenatbib]{mnras}

\usepackage{mathptmx}

\usepackage[T1]{fontenc}

\DeclareRobustCommand{\VAN}[3]{#2}
\let\VANthebibliography\thebibliography
\def\thebibliography{\DeclareRobustCommand{\VAN}[3]{##3}\VANthebibliography}

\usepackage{graphicx}	
\usepackage{amsmath}	
\usepackage{amssymb}	
\usepackage[normalem]{ulem}
\usepackage{color}

\usepackage[dvipsnames]{xcolor}

\def\kms{{\rm\,km\,s^{-1}}}

\title[Centre of mass properties]{Exploring the centre of mass properties of LG-like galaxies}

\author[Jean-Baptiste Salomon et al.]{
Jean-Baptiste Salomon,$^{1,2,3}$\thanks{E-mail: jean-baptiste.salomon@utinam.cnrs.fr}
Noam Libeskind$^{2}$ and
Yehuda Hoffman$^{1}$
\\
$^{1}$Racah Institute of Physics, Hebrew University, Jerusalem 91904, Israel\\
$^{2}$Leibniz-Institut für Astrophysik Potsdam, An der Sternwarte 16, D-14482 Potsdam, Germany\\
$^{3}$Institut UTINAM, CNRS UMR6213, Univ. Bourgogne Franche-Comt\'e, OSU THETA,
   Observatoire de Besançon, BP 1615, 25010 Besan\c{c}on C\'edex, France\\
}

\date{Accepted 2023 May 23. Received 2023 May 15; in original form 2023 January 07}

\pubyear{2023}

\begin{document}
\label{firstpage}
\pagerange{\pageref{firstpage}--\pageref{lastpage}}
\maketitle

\begin{abstract}
From high resolution cosmological simulations of the Local Group in realistic environment, namely HESTIA simulations, we study the position and kinematic deviations that may arise between the disc of a Milky Way (or Andromeda)-like galaxy and its halo. We focus on the 3-dimensional analysis of the centres of mass (COM). The study presents two parts. We first consider individual particles to track down the very nature and amplitude of the physical deviations of the COM with respect to the distance from the disc centre. Dark matter dominates the behaviour of the COM of all particles at all distances. But the total COM is also very close to the COM of stars. In the absence of a significant merger, the velocity offsets are marginal (10$\kms$) but the positional shifts can be important compared to the disc characteristics (> 10 kpc). In the event of a massive accretion, discrepancies are of the same order as the recent finding for the MW under the Magellanic Clouds influence. In a second part, the accent is put on the study of various populations of subhaloes and satellites. We show that satellites properly represent the entire subhalo population. There exists strong mismatch in phase space between the satellites' COM and the host disc. Moreover, the results are highly inhomogeneous between the simulations, and thus between the accretion histories. Finally, we point out that these shifts are mainly due to a few of the most massive objects.
\end{abstract}




\section{Introduction}
A thorough knowledge of the most basic parameters of a galaxy, namely the position and velocity of its centre, is essential to understand how galaxies move with respect to each other. In the $\Lambda$CDM framework, the classic model for a massive spiral galaxy is that of an equilibrium disc in a stationary state at the very centre of its (possibly) triaxial halo. Objects such as dwarf galaxies orbit inside the halo and around the central galaxy. This simple picture is often adopted, especially in the Local Group (LG), because it allows cosmologists to model complicated systems. 
For example, if the Andromeda galaxy (M31) and the Milky Way (MW) - the two main galaxies of the LG - are treated as point particles, the so-called ``timing argument'' allows for an estimation of the LG mass assuming some cosmological parameters (e.g. \citealt{Penarrubia14}). 
Such hypotheses have enabled scientists to make great advances; however it is also acknowledged that these are simplifying assumptions that only provide an approximate description of nature.

Upon closer examination, a number of faults can be identified. First of all the central disc can be intrinsically perturbed, for example by spiral arms or bar buckling instabilities \citep{Faure14,Debattista14,Monari16,Khoperskov19}. Secondly, the structure of the disc can also be perturbed by external interference like a dwarf galaxy being accreted or a satellite encounter \citep{Gomez13,Widrow14,Laporte18,Chequers18}. Finally, the population of satellites is not necessarily relaxed and the mass of the satellites is not always negligible. Both the disc and the halo can then undergo deformations and warpings as for example the MW under the effect of the Magellanic clouds \citep{Weinberg98,GaravitoCamargo19,Conroy21}. Hence, the halo and the disc can have different dynamics \citep{Petersen21}.

Consequently, when the galaxy is perturbed and especially in the case where much mass is held in satellites, the system goes out of stationary equilibrium \citep{Erkal21}. In this situation, the centre of the disc of the host galaxy does not necessarily coincide with the centre of the dynamical system as a whole (i.e. the halo plus satellites). Although conflating these two centres is convenient, it risks moving away from the true physical nature of the system. For example, the calculations of satellite orbits can be skewed when the centre of the host galaxy is artificially locked \citep{White83,Gomez15}.
Also, the local dark matter density can be overestimated by 20\% if the false assumption that the Galaxy is in equilibrium is taken \citep{Banik17,Haines19,Salomon20}. The interpretation of the content and dynamics of the halo is then biased.
The mass of the MW enclosed in a larger radius, from 100 to 200 kpc, can also be largely overestimated in an equilibrium scenario - from about 15\% up to 50\% \citep{Erkal20, CorreaMagnus22}. Last but not least, the application of the timing argument model to the LG shows a mass of a few tens of percent lower when the MW is considered out of equilibrium than when it is considered in equilibrium \citep{Benisty22,Chamberlain22}.

These simplifying assumptions may induce non-uniform effects on the different methods employed for measuring the PM of M31 \citep{vdM08, Sohn12, Salomon16, vdM19, Salomon21}. Indeed, current values of the relative transverse velocity of the M31 galaxy with respect to the MW show great discrepancy (see Figure 6 in \citealt{Salomon21}). This is especially true when comparing the values derived by 'direct' methods with those obtained by 'indirect' methods. Direct methods  rely on the study of the proper motions of individual stars identified as belonging to the M31's disc to derive an overall motion of the galaxy \citep{Sohn12, vdM19, Salomon21}. Indirect methods, on the other hand, study the ensemble motion of the satellites of M31, under the assumption that the satellites, being embedded in the main halo, follow the same overall mean velocity as their host galaxy \citep{vdM08, Salomon16}.
The different values of transverse velocities imply different trajectories: towards the south-east for the direct  methods and towards the north-west for the indirect methods. Of course, one could argue that a reasonable value is the median, thus favouring a purely radial orbit.
This is arguable especially in view of the large uncertainties with both types of methods. But another way of approaching the problem is to consider that we are comparing results of two fundamentally different things. For example the disc itself, but also the population of satellites and thus the entire halo may not be in equilibrium. Therefore those different components will not have the same position and velocity of barycentre.

Aware of these pitfalls, the community is making great progress in the study of non-equilibrium models. However, the efforts are mainly concentrated on the most important disturber of the MW, the Large Magellanic Cloud (LMC). The same is true for N-body simulations which try to reproduce the LG. If these efforts are crucial to improve our knowledge, it is also necessary to study in a more broad-based approach the impact of the whole content of a halo and the decoupling appearing with its central disc, in a cosmological context.

Hence, the hypothesis we propose to explore in this article is to use constrained cosmological simulations of the LG to qualitatively evaluate the extent to which the central baryonic disc, the satellites and the halo are offset in terms of their position and kinematics. The paper is organised as follows. In Section~\ref{sect:simulations}, we present the frame of work which are the simulated data and benchmarks used in the study. 
Then, we undertake a comprehensive and extensive study of the evolution of the positions and velocities of the centres of mass (COM) of the different types of particles for each host galaxy in Section~\ref{sect:COMpart}. More specifically, we compare the phase-space configuration of the COM of the disc with that of its halo content with respect to the distance to the centre. 
In Section~\ref{sect:COMshalo}, we study the COM of subhaloes populations while in Section~\ref{sect:COMsat}, we focus on satellites. 
Then in Section~\ref{sect:COMmassive}, we statistically investigate the impacts of massive satellites. Finally, we summarise and conclude in Section~\ref{sect:COMccl}.

\section{Simulations}\label{sect:simulations}

\subsection{Data}

The High-resolutions Environmental Simulations of The Immediate Area (HESTIA) project \citep{Libeskind20} is a suite of magneto-hydrodynamical cosmological simulations run with the moving mesh cosmological simulation code Arepo \citep{Springel10} that employs the Auriga model \citep{Grand17} for star formation and feedback. We refer the reader to \citep{Libeskind20} for details regarding the simulations and only highlight the most salient points here. The HESTIA simulation are constrained simulations of our local environment that use as input the local cosmography as described by the peculiar velocity field \citep{Sorce14}. The initial conditions have been carefully selected in order to  reproduce - at redshift zero - a LG similar to the observations. Hundreds of simulations have been run and those which most closely reproduce the LG, and its environment are kept for further study including high resolution simulations.

The three high resolution simulations used in this study, named~ 09\_18, 17\_11 and 37\_11,  thus have a similar cosmological environment. Those LGs contain two giant spiral galaxies of mass equivalent to that of the MW and the M31 galaxy. 
They dominate their immediate environment which means that there are no other large galaxies near them. And a cluster, of the same mass as Virgo ($\gtrsim$10$^{14}\rm{M}_{\odot}$), 
is located at the equivalent distance ($\sim$17 Mpc).  
Finally, the LG galaxies are separated by a distance of around 0.7 Mpc and are approaching each other or are close to doing so. The properties of these three simulations relevant to this work are summarised in Table~\ref{tab:simus}. 

\begin{table}
	\centering
	\caption{Main properties of the three HESTIA high resolution simulations containing a Local Group analog at z=0. The first column presents the identifiers of the simulations. The masses M$_{200}$ (mass enclosed within the radius within which the mean density is 200 times the critical density) of the two most massive haloes in the simulated Local Group is given in the second (for the Andromeda galaxy) and third columns (for the Milky Way). The fourth column is the separation between the centre of these two haloes. The relative radial and tangential velocities between the two main haloes are given in the fifth and sixth columns.}
	\label{tab:simus}
	\begin{tabular}{lccccc} 
		\hline
		Name    & M$_{\rm M31}$ & M$_{\rm MW}$  & d     & v$_{\rm r}$   & v$_{\rm t}$\\
		        & 10$^{12}$ M$_{\odot}$ & 10$^{12}$ M$_{\odot}$ & kpc & $\kms$ & $\kms$ \\
		\hline
		09\_18  & 2.13          & 1.94          & 866   & -74.0         & 54.0 \\
		17\_11  & 2.30          & 1.96          & 675   & -102.2        & 137  \\
		37\_11  & 1.09          & 1.04          & 850   & 8.86         & 71.1 \\
		\hline
	\end{tabular}
\end{table}

\begin{figure*}
    \centering
        \begin{tabular}{ccc}
    \includegraphics[width=5.6cm] {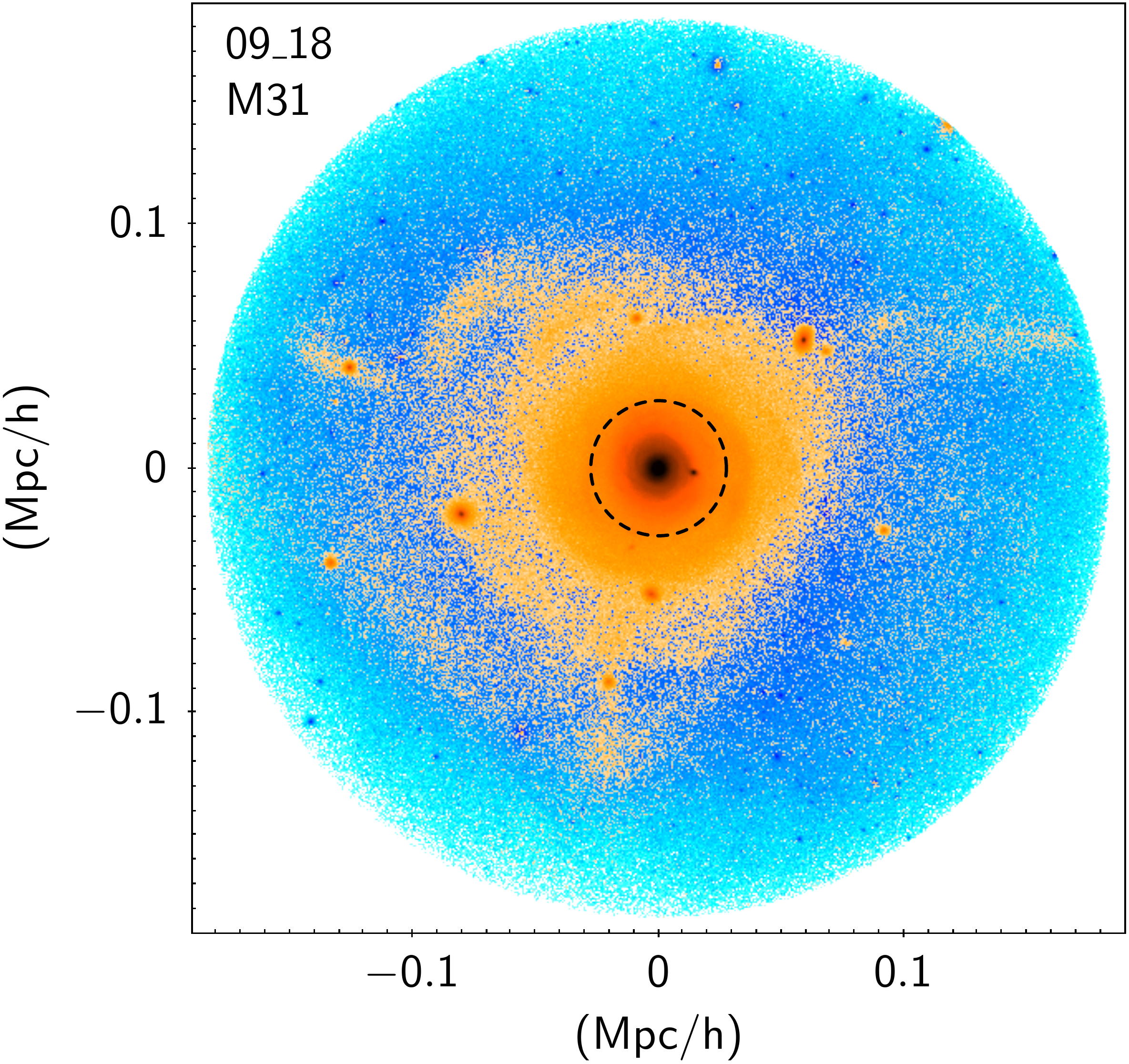} & 
    \includegraphics[width=5.6cm] {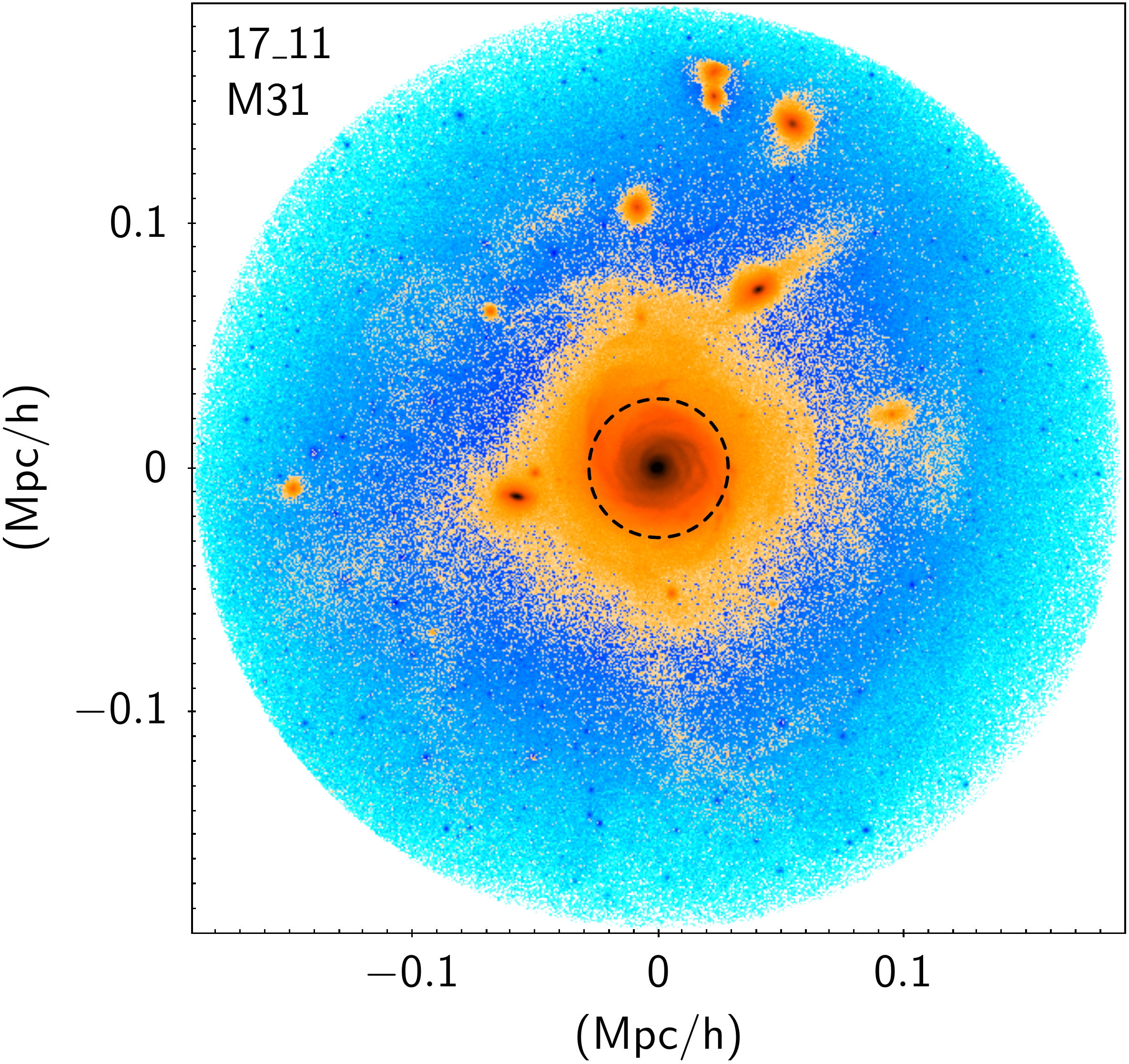} & 
    \includegraphics[width=5.6cm] {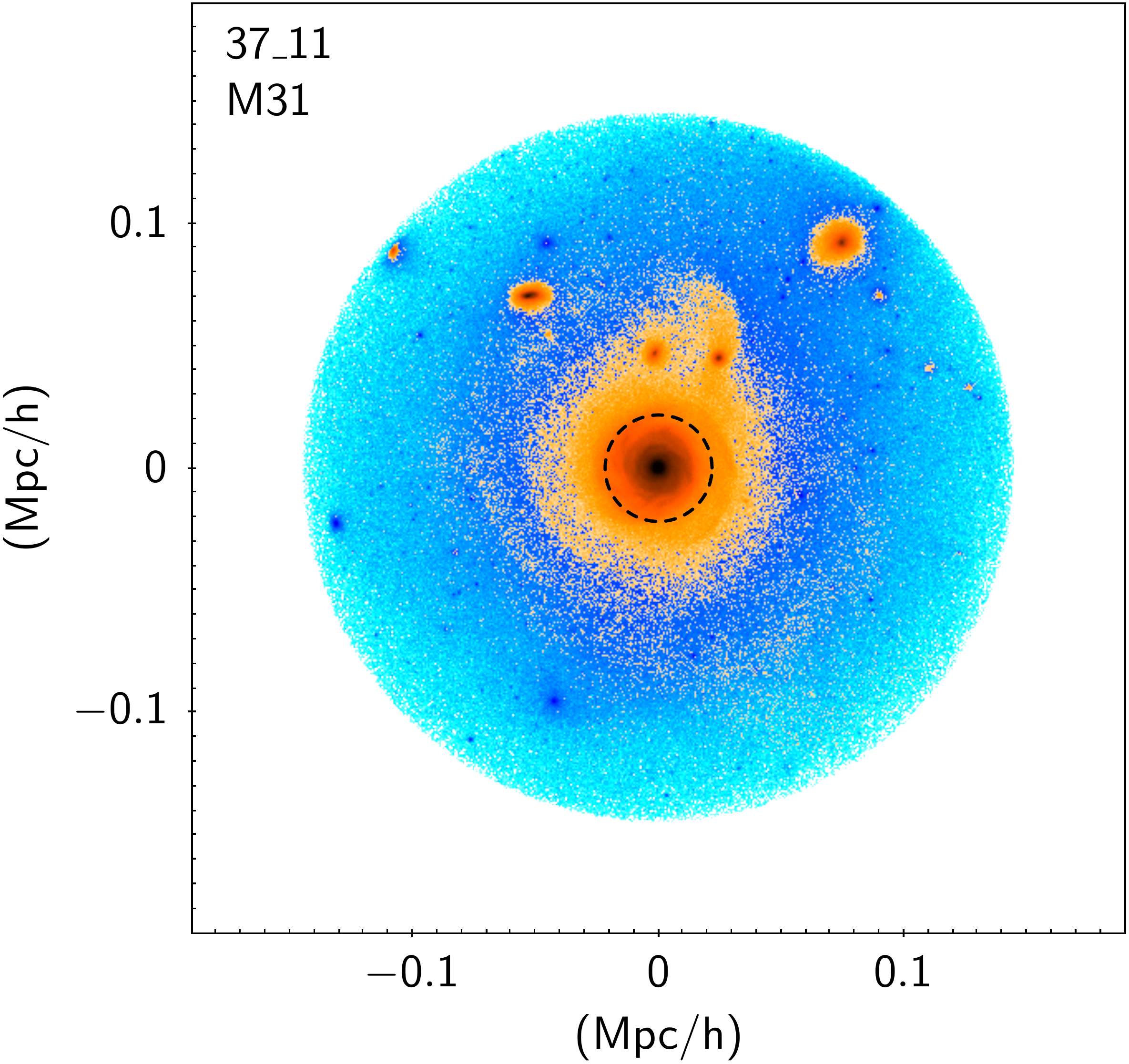} \\
    \includegraphics[width=5.6cm] {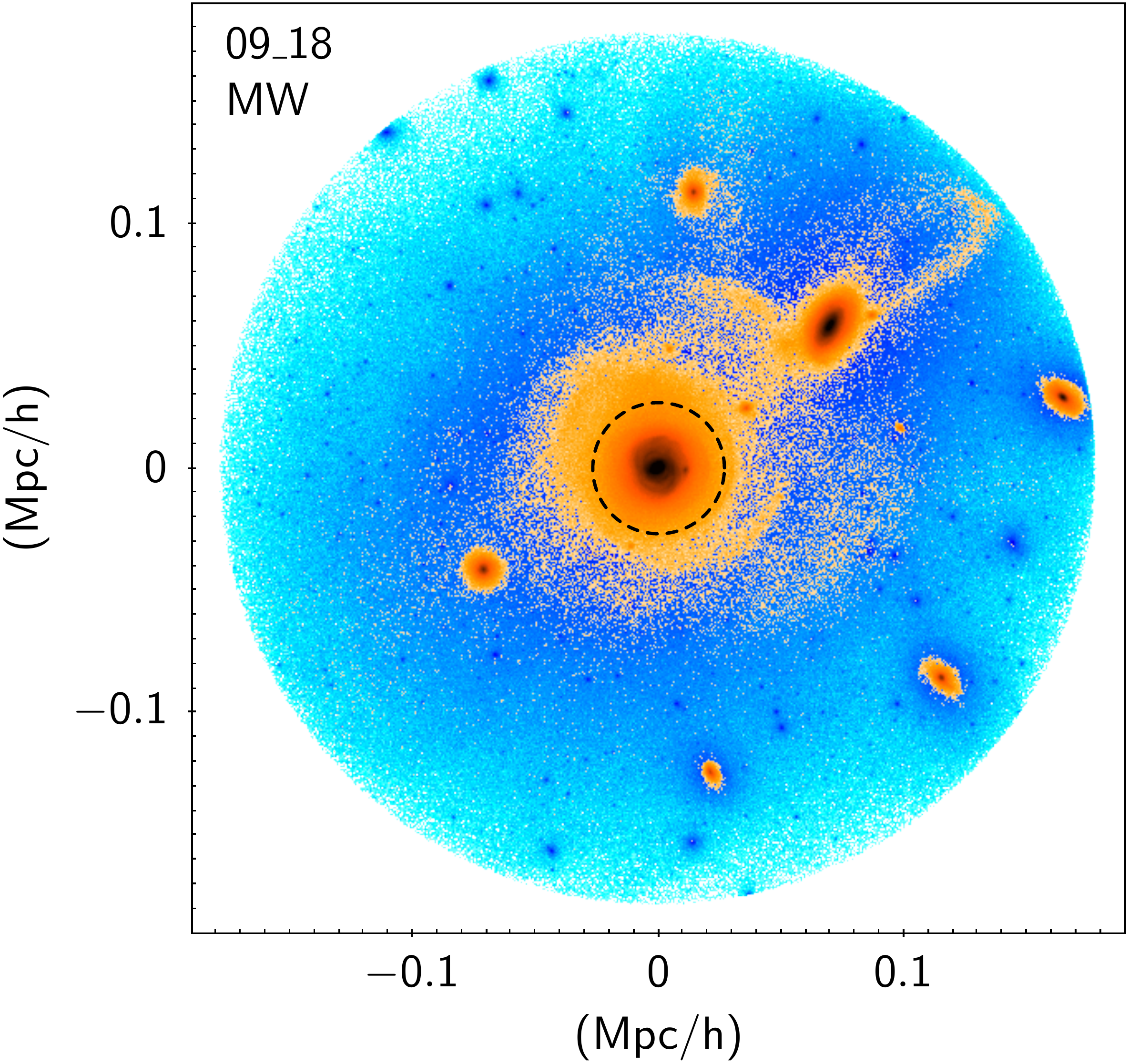} & 
    \includegraphics[width=5.6cm] {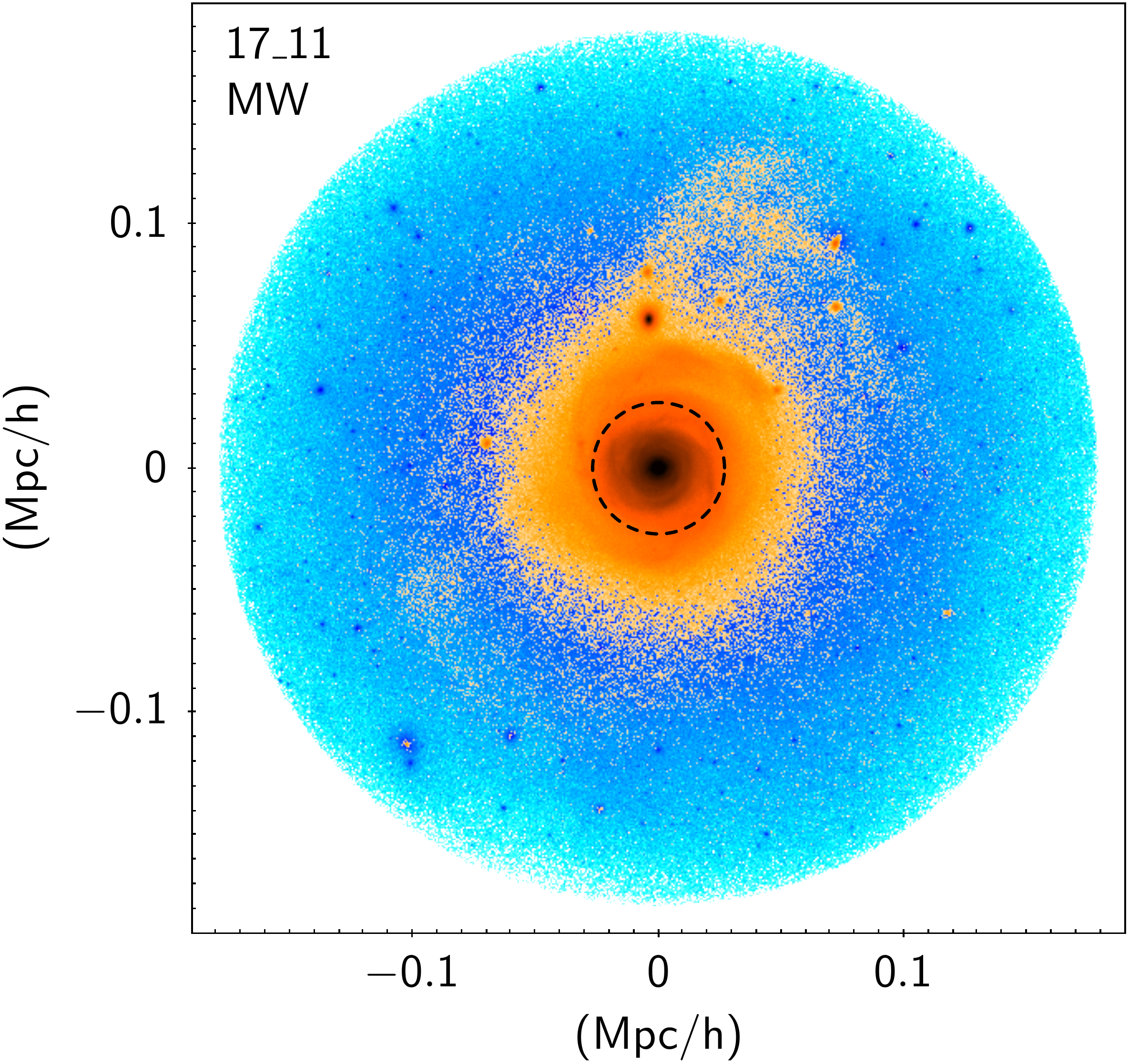} & 
    \includegraphics[width=5.6cm] {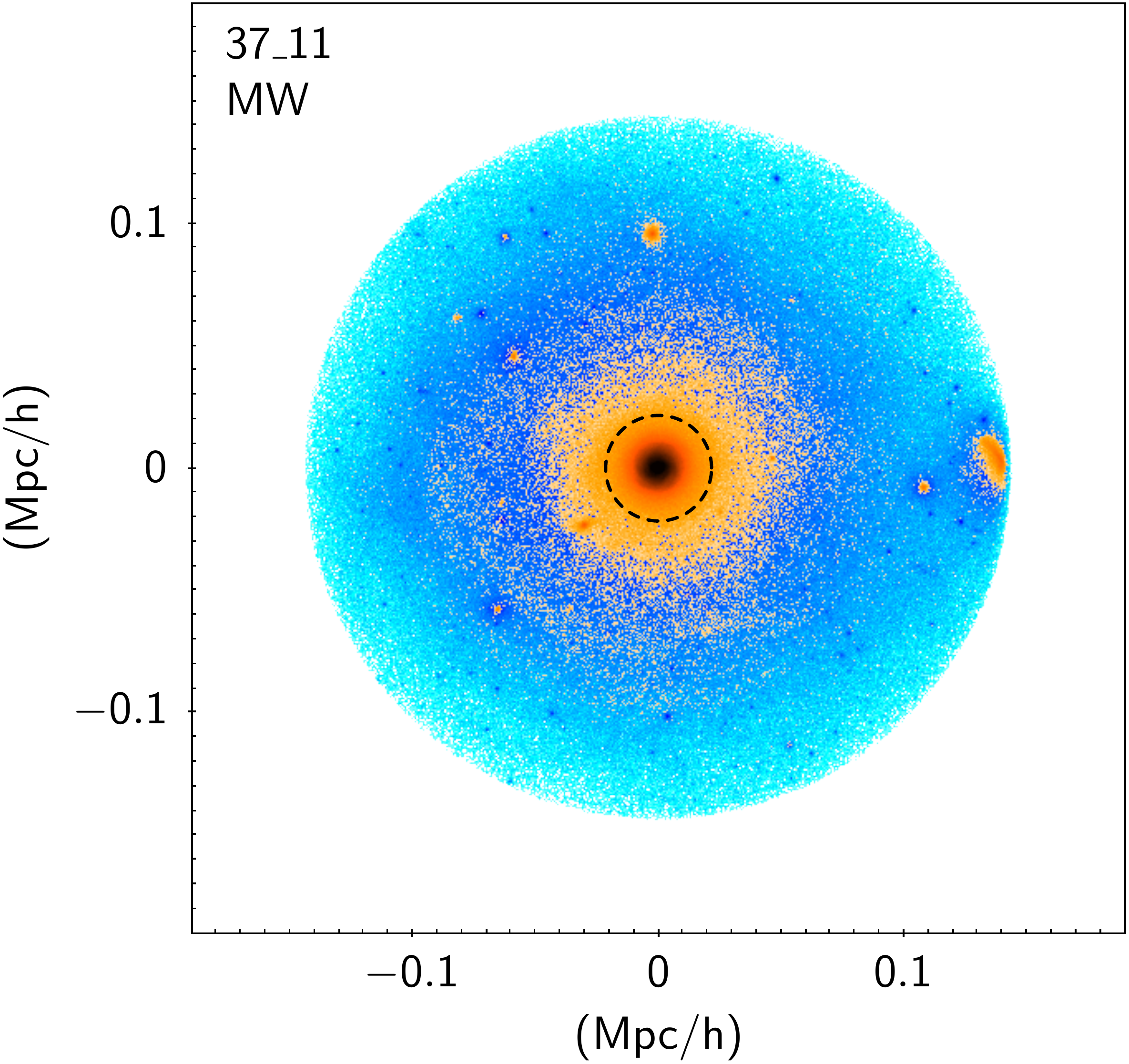} \\
   \end{tabular}
\caption{Face on projection at z=0 of the six host galaxies from the three high resolution HESTIA simulations of the LG analogues, 09\_18, 17\_11 and 37\_11 from left to right. Figures are centred on the disc centre of mass as derived in section~\ref{discCentre}. The names M31 (top row) and MW (bottom row) distinguish the more and less massive host in each of the simulated LG. Blue gradient colour traces density of dark matter particles from low (light blue) to high (dark blue) densities. Orange gradient colour traces density of star particles from low (light orange) to high (dark orange) densities. Only particles within R$_{200}$ of the host galaxy and identified as gravitationally bounded to it are represented. The dotted black circle delimits the radius of $0.15 \times \rm{R}_{200}$ considered in this study within which the particles belonging to the host have been used to calculate the central disc properties.}
    \label{img_simu_m31}
\end{figure*}

In Section~\ref{sect:COMpart}, we study the behaviour of all the particles belonging to the host galaxy. There are four kinds of particles (or equivalent cells) in the simulations: gas, dark matter, star and black hole. The spatial resolution is 220 pc with a mass resolution of 1.5 $\times$ 10$^{5}$M$_{\odot}$ for dark matter and 2.2 $\times$ 10$^{4}$M$_{\odot}$ for the gas. Since we have three high-resolution LG systems, there are six host galaxies similar to M31 (see Figures~\ref{img_simu_m31}). Note that the nomenclature ``M31'' and ``MW'' used throughout the HESTIA project is somewhat arbitrary: the two LG members have roughly the same mass, as such the more massive one was termed M31.

%
%

Only the final $z=0$ snapshot of each simulation is used. The properties of the haloes, subhaloes as well as the identification of the particles composing them have been identified with the Adaptive Mesh Investigations of Galaxy Assembly (Amiga) Halo Finder (AHF,  \citealt{Gill04,Knollmann09}). In brief, AHF lays a hierarchy of grids, refining more and more in order to identify iso-density contours. The halo finder defines a halo centre as the position of the densest cell of the highest refinement level. The halo boundary is determined as R$_{200}$, the radius within which the mean density is 200 times the critical density.

\subsection{Disc centre of mass}\label{discCentre}

The collections of particles bound to each of the six host galaxies within R$_{200}$ are identified by AHF.
To define the galactic disc for a given host, we first remove the particles picked out as belonging to subhaloes or satellites creating a kind of ``Swiss cheese'' topology. The remaining particles are called ``the host sample''. 
All the particles or equivalent cells (star, dark matter, black hole and gas) from the host sample contained in a sphere centred on the AHF centre of the main halo and of radius $0.15 \times \rm{R}_{200}$ are then considered.
The barycentre of this set of particles is calculated. Then, a new sample is built from the host sample again, defined as the ensemble of particles contained in a sphere centred on the new barycentre and again of radius $0.15 \times \rm{R}_{200}$. The calculation is reiterated until convergence - about ten times. Eventually, the position ($\vec{\rm{x}}_{\rm{c}}$) and the velocity ($\vec{\rm{v}}_{\rm{c}}$) of the centre of mass (COM)\footnote{We use the terms barycentre and centre of mass interchangeably throughout this paper} of the host galaxy is obtained.

We then calculate the angular momentum of this host sample. A new Cartesian coordinate referential frame centred on $\vec{\rm{x}}_{\rm{c}}$ is adopted for the simulations where the z-axis is aligned with the angular momentum vector. It allows to compute the velocity space (v$_{\rm{r}}$,v$_{\rm{\theta}}$,v${\rm{_z}}$) centred on $\vec{\rm{v}}_{\rm{c}}$. In each of the simulations, the distribution of v$_{\rm{r}}$ and v$_{\rm{z}}$ is a near symmetric normal distribution centred on 0 while the distribution of v$_{\rm{\theta}}$ exhibits a Poisson shape with maximum values between 200 and 250 $\kms$. Hence, the velocity spaces clearly illustrate a rotating structure for host sample's stars in all simulations. Moreover, the star particles dominate the other components in the very inner part in term of the mass budget. These properties allow us to claim that the centre ($\vec{\rm{x}}_{\rm{c}}$,$\vec{\rm{v}}_{\rm{c}}$) found is coincident with the centre of the host disc.

The value of 15\% of R$_{200}$ is motivated by the observed MW since the disc is around 15 kpc (see e.g. \citealt{Robin92}) and its radius R$_{200}$ is around 220 $\pm$ 60 kpc (see e.g. \citealt{McMillan11}). Selecting particles within a radius of about twice that of the MW's disc is sufficiently large to be ensure the entire disc is targeted.
The results shown here are more or less indifferent to reasonable assumptions regarding the disc size in these simulations.
Specifically, changing the disc size by making it larger or smaller by 2\% of R$_{200}$ (i.e. approximately $\pm$ 5 kpc) changes the centre by less than the softening length (220 pc). The velocity is modified by a maximum of 1.3 $\kms$. Larger disc radius modifications ($\pm$ 10\% of R$_{200}$, approximately $\pm$ 25 kpc) lead to a variation of the centre position of not more than four times the smoothing length (880 pc) and of the centre velocity of at most 6$\kms$. To remain conservative, we will neglect in the following any positional shifts below 1 kpc and the velocity offsets smaller than 6$\kms$. We also point out that throughout the rest of the paper we will be comparing peculiar velocities, as the differences in distances we will encounter are negligible compared to the typical cosmological distances of the Mpc.

In Figures~\ref{img_simu_m31} we present the face on projections (perpendicular to the angular momentum vector previously calculated) of each simulated MW and M31 under consideration. Each galactic disc is demarcated by a dotted black circle. The final two vectors $\vec{\rm{x}}_{\rm{c}}$ and $\vec{\rm{v}}_{\rm{c}}$ will be used for the rest of the study as references for position and velocity comparisons respectively.

\section{Centre of mass deviation for particles}\label{sect:COMpart}

\subsection{Positional offsets at the particle level}
\begin{figure*}
    \centering
        \begin{tabular}{ccc}
    \includegraphics[width=5.6cm]{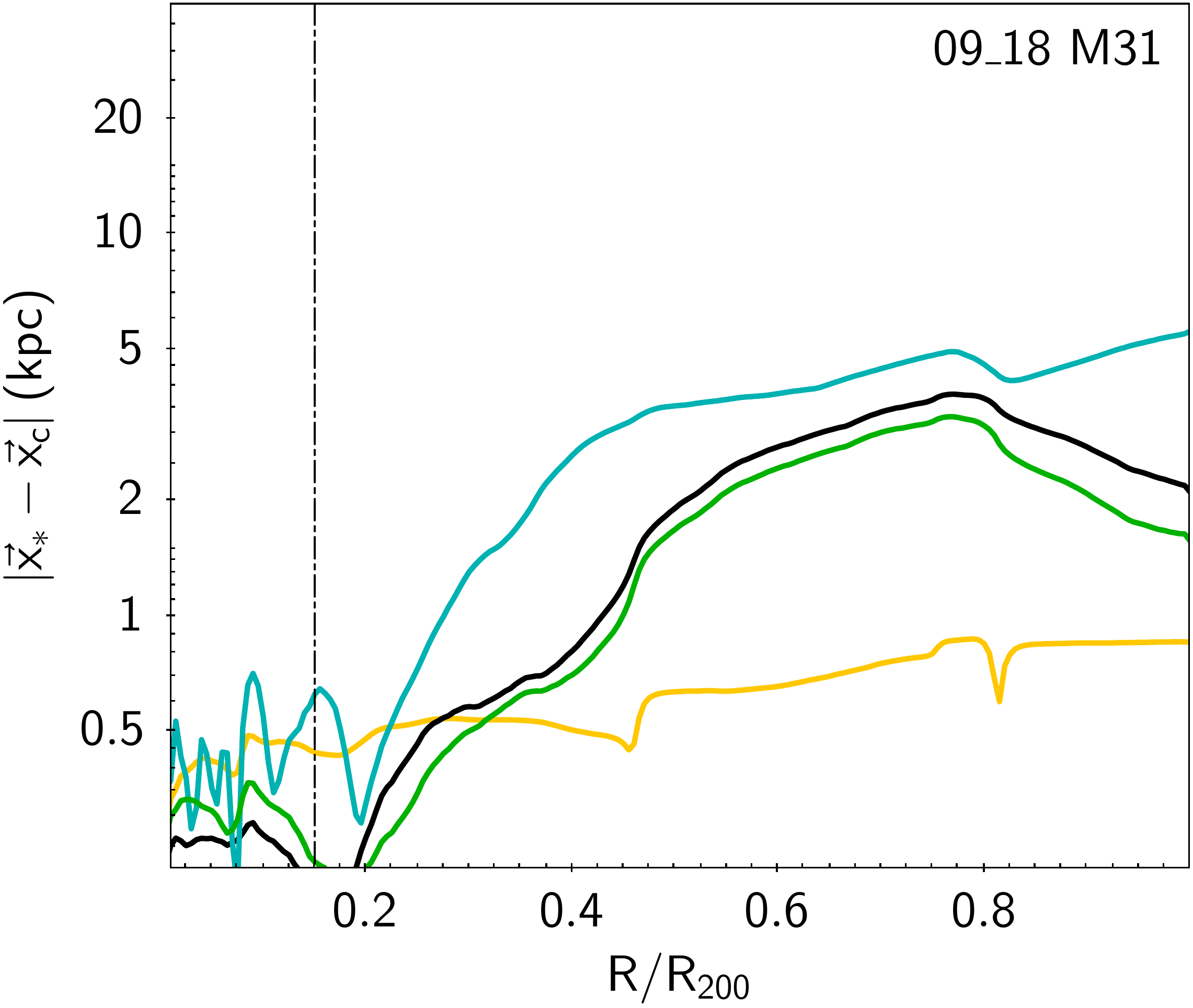} &
    \includegraphics[width=5.6cm]{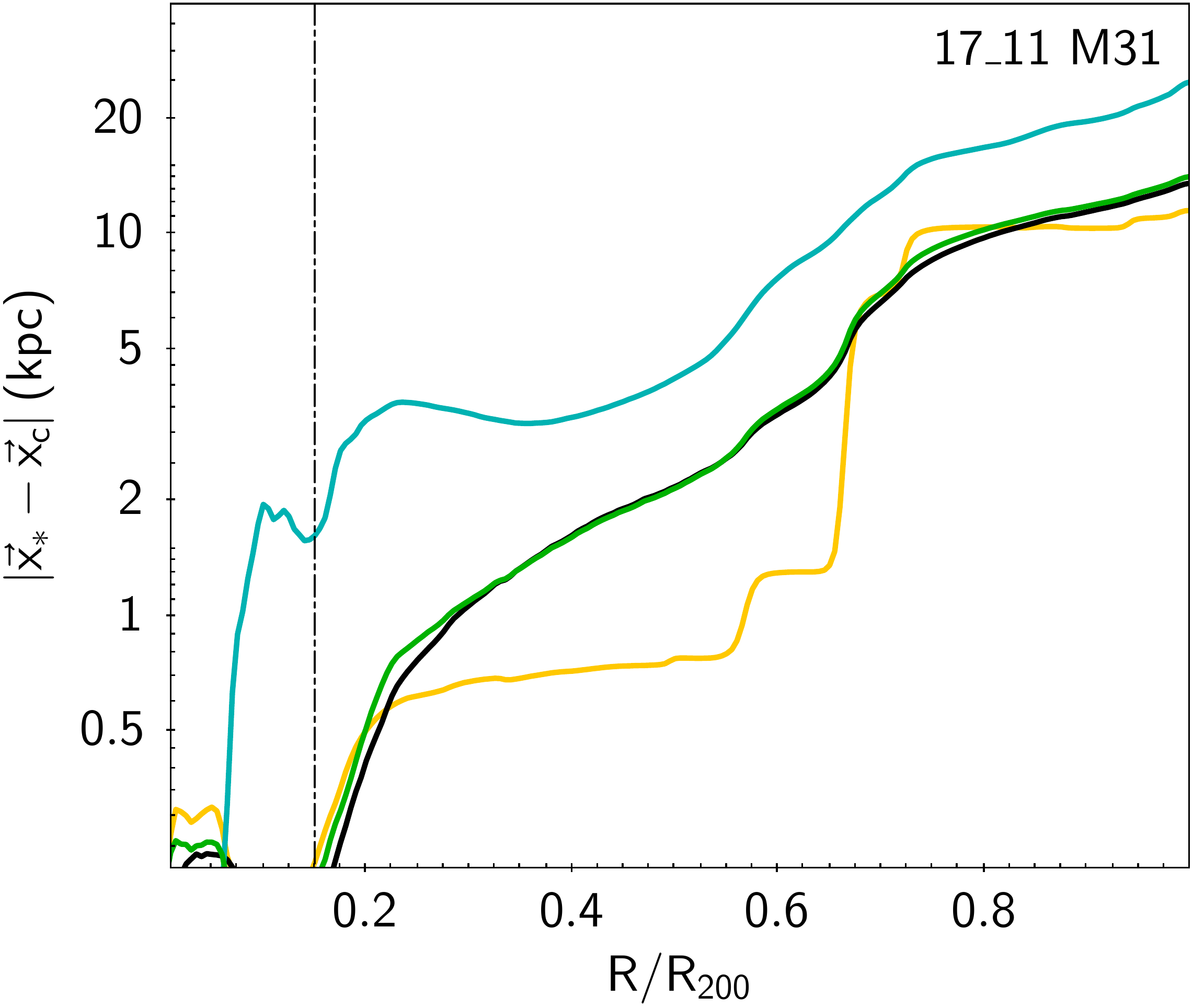} & \includegraphics[width=5.6cm]{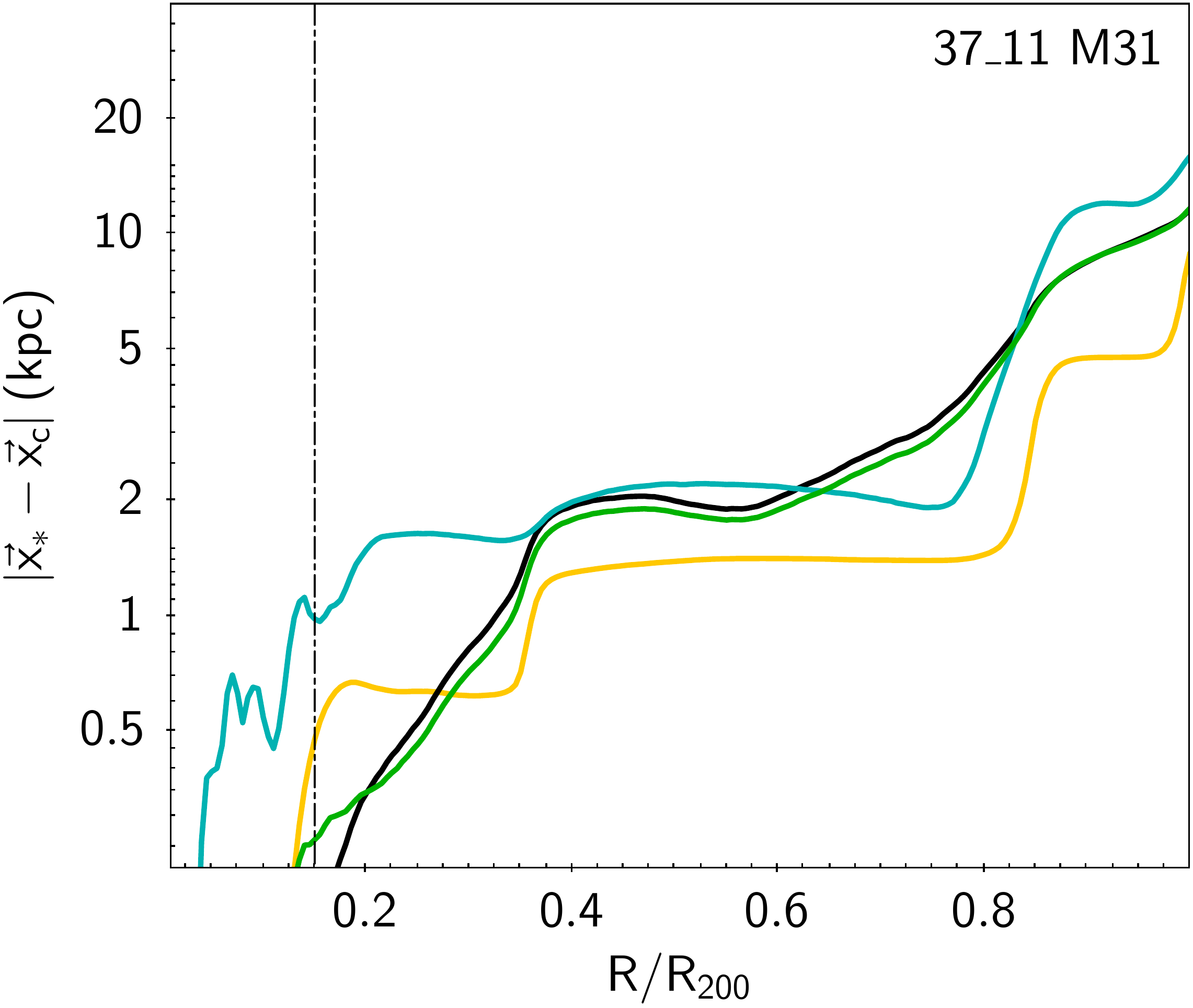}  \\
    \includegraphics[width=5.6cm]{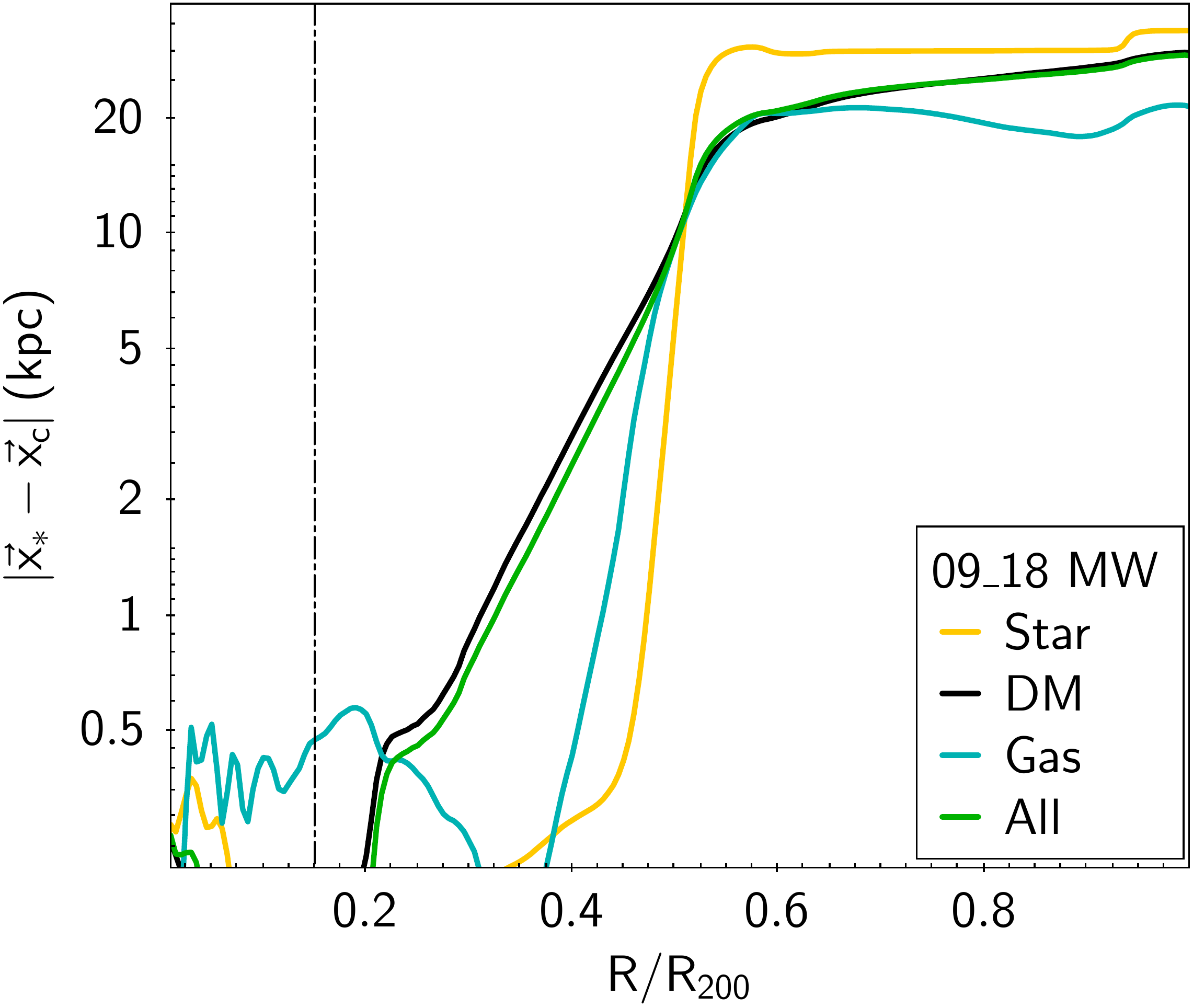} &
    \includegraphics[width=5.6cm]{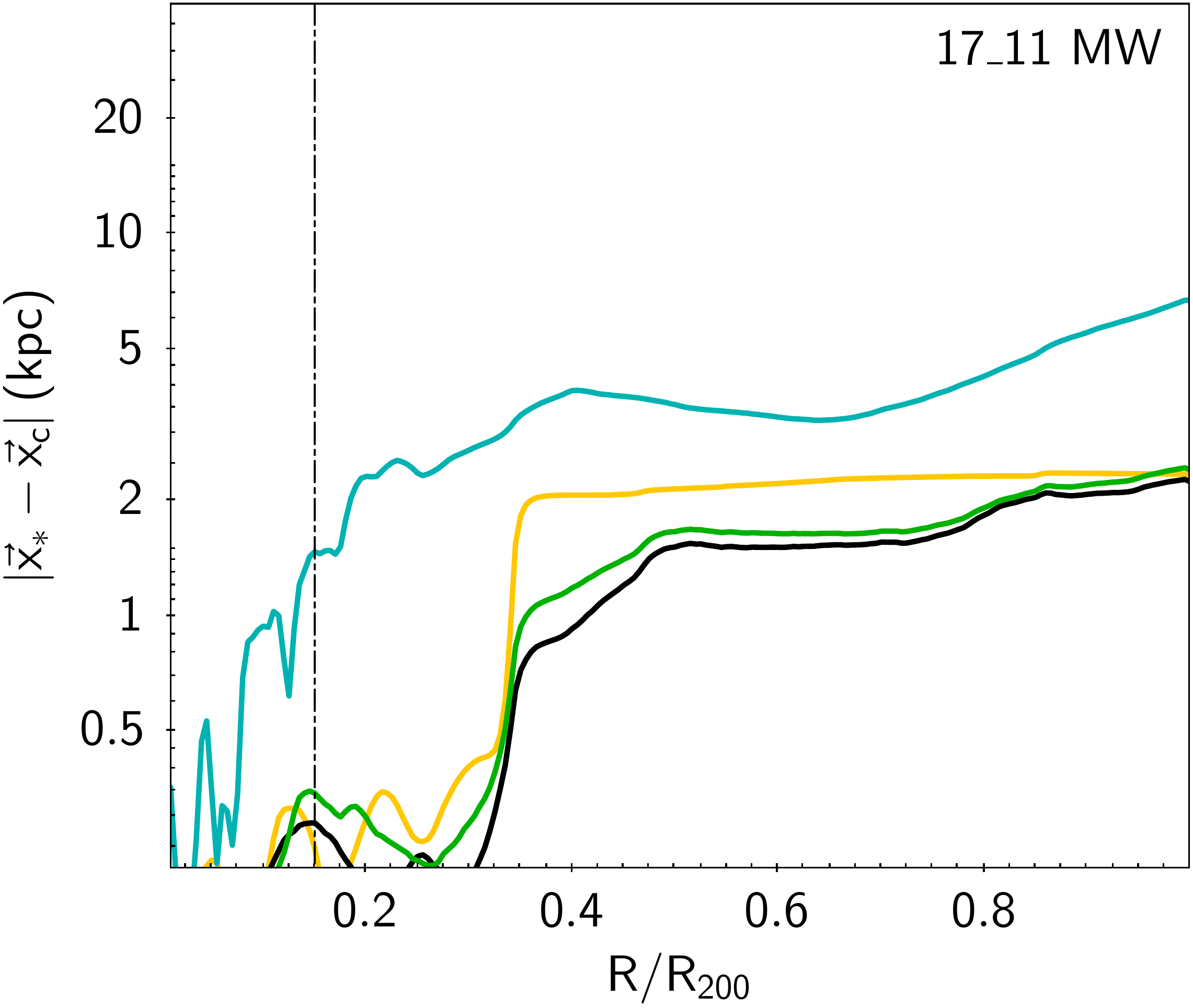} & 
    \includegraphics[width=5.6cm]{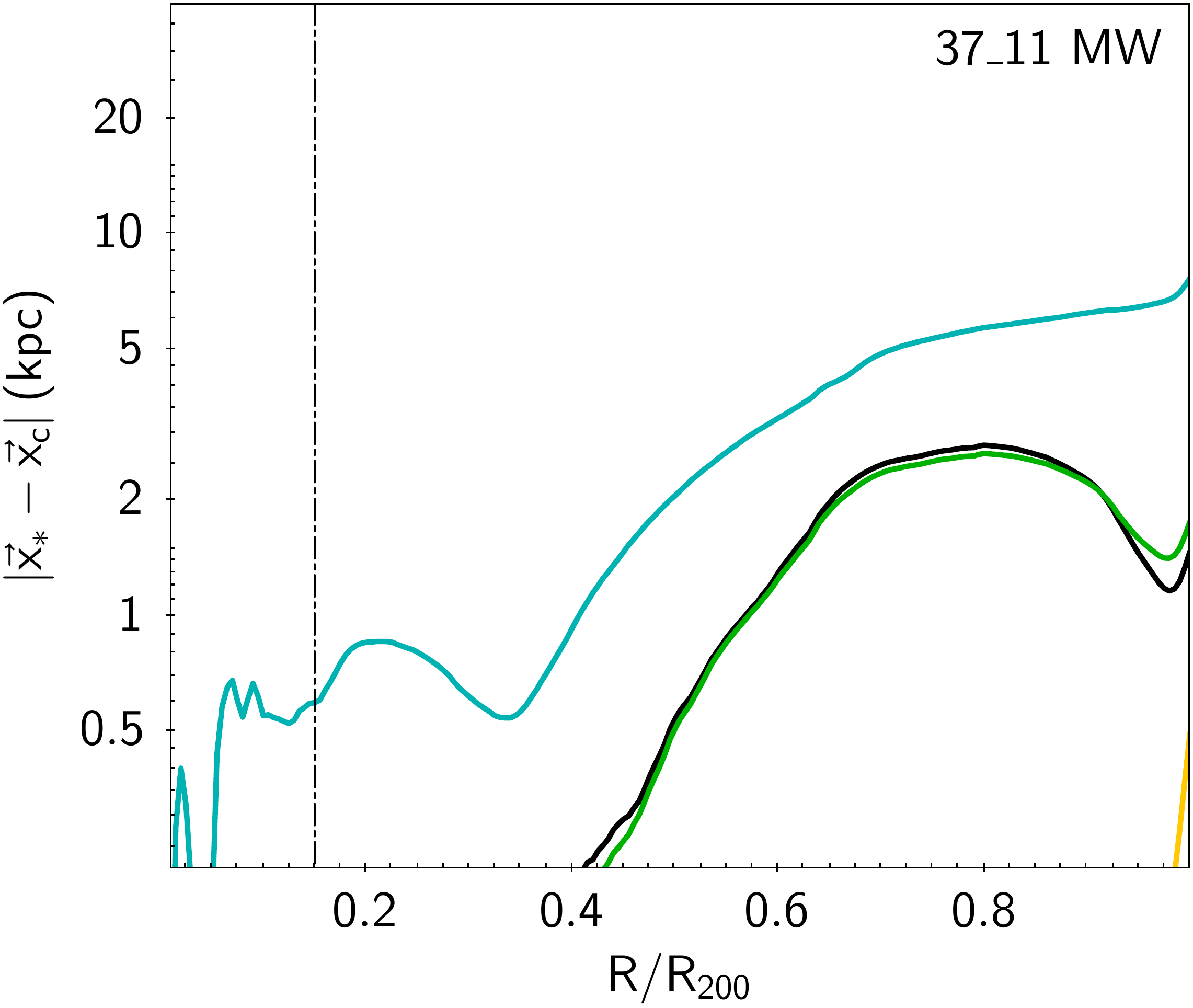} \\
   \end{tabular}
\caption{For the six host galaxies in the high resolution HESTIA simulations, absolute differences in position between the disc centre and the centres of mass of different types of particles (or equivalent cells) contained within a sphere of radius R, for stars in yellow, dark matter in black, gas in cyan and all components in green. 
The maximum radius corresponds to R$_{200}$. The vertical dotted line represents the radius of the sphere ($0.15 \times \rm{R}_{200}$) taken into account to calculate the disc centre of mass position and velocity. We set the minimum value of the y-axis equal to 220 pc, the smoothing length. Note that given the uncertainties and resolution effects, any offset of less than 1 kpc should be treated with caution (see Section \ref{discCentre}).}
    \label{diffx_part}
\end{figure*}

We begin our investigation by examining the deviations in mean position and velocity between the disc centre (namely $\vec{\rm{x}}_{\rm{c}}$ and $\vec{\rm{v}}_{\rm{c}}$), and three different components of the simulation (i.e. star particles, dark matter particles, and gas cells) as a function of the distance. We recall that $\vec{\rm{x}}_{\rm{c}}$ and $\vec{\rm{v}}_{\rm{c}}$ are computed from all components in the iterative way described above. At a distance $0 \le \rm{R} \le \rm{R}_{200}$, the particles identified as being bound to the main halo and centred on the ${\rm{x}}_{\rm{c}}$ position are considered. At a given radius, the COM position of each particle type as well as that of all particles in addition to black holes is derived and compared to the COM position of the disc (see Figure~\ref{diffx_part}). In a similar way, the velocities of the different COMs are calculated at each radius and compared with the COM velocity of the disc (see Figure~\ref{diffv_part}).
%

%

Concerning the offsets in position, several points can be noted. First, in the central part, corresponding to the disc, the deviations for the stellar and dark matter particles are minimal, well below 1 kpc. Only the COM of the gas cells can slightly deviate from the COM of the disc, highlighting the non-homogeneity of the gas distribution in and around a spiral galaxy. The impact on the behaviour of the whole system is however negligible because the gas component is diffuse and not very massive compared to the dark matter component or the stellar component. 

Secondly, the general behaviour of the deviations is dominated by that of the dark matter particles at any radius. The green curve in Figure~\ref{diffx_part} (all particles) is very close to the dark curve (dark matter particles). This feature is expected as the dark component gravitationally dominates the halo.

In addition, the stellar component is not strictly correlated with the dark component. The star COM undergoes steeper  variations and fluctuations in position as a function of radius. By contrast, the dark component, and hence the COM of the whole system, go through smoother deviations.
This reflects both the small fraction of stellar particles present outside the disc of the host galaxy and also the intrinsic nature of dark matter particles themselves which have a weak interaction with their environment, and consequently a less sharp distribution.

The positional shifts between the disc and the COMs of the system are often only increasing as the radius of the sphere containing the particles considered increases.
We might have expected not to have this effect. Indeed, with a relatively homogeneous system, considering a larger volume will smooth out the bumps and decrease the deviations.
The differences at R$_{200}$ are at least an order of magnitude larger than the differences between the edge of the disc and its centre. The deviations are particularly marked when a massive satellite is present at a given radius (see for example simulation 09\_18 of the MW). This results in large diversity of positional shifts at R$_{200}$ between the disc COM and the host halo system COM, spanning the values [1.5, 1.6, 2, 11.5, 14, 30] kpc. Some of these deviations are very important with respect to the typical size of a spiral galaxy, or even to the solar radius for example which is about 8 kpc. 
Thus, these misalignments, arising from the merger history or directly from the presence of a satellite, cannot be neglected in the study of the dynamics of a typical MW (or M31) galactic system at the scale of its halo.

\subsection{Velocity shifts at the particle level}

\begin{figure*}
    \centering
        \begin{tabular}{ccc}
    \includegraphics[width=5.6cm]{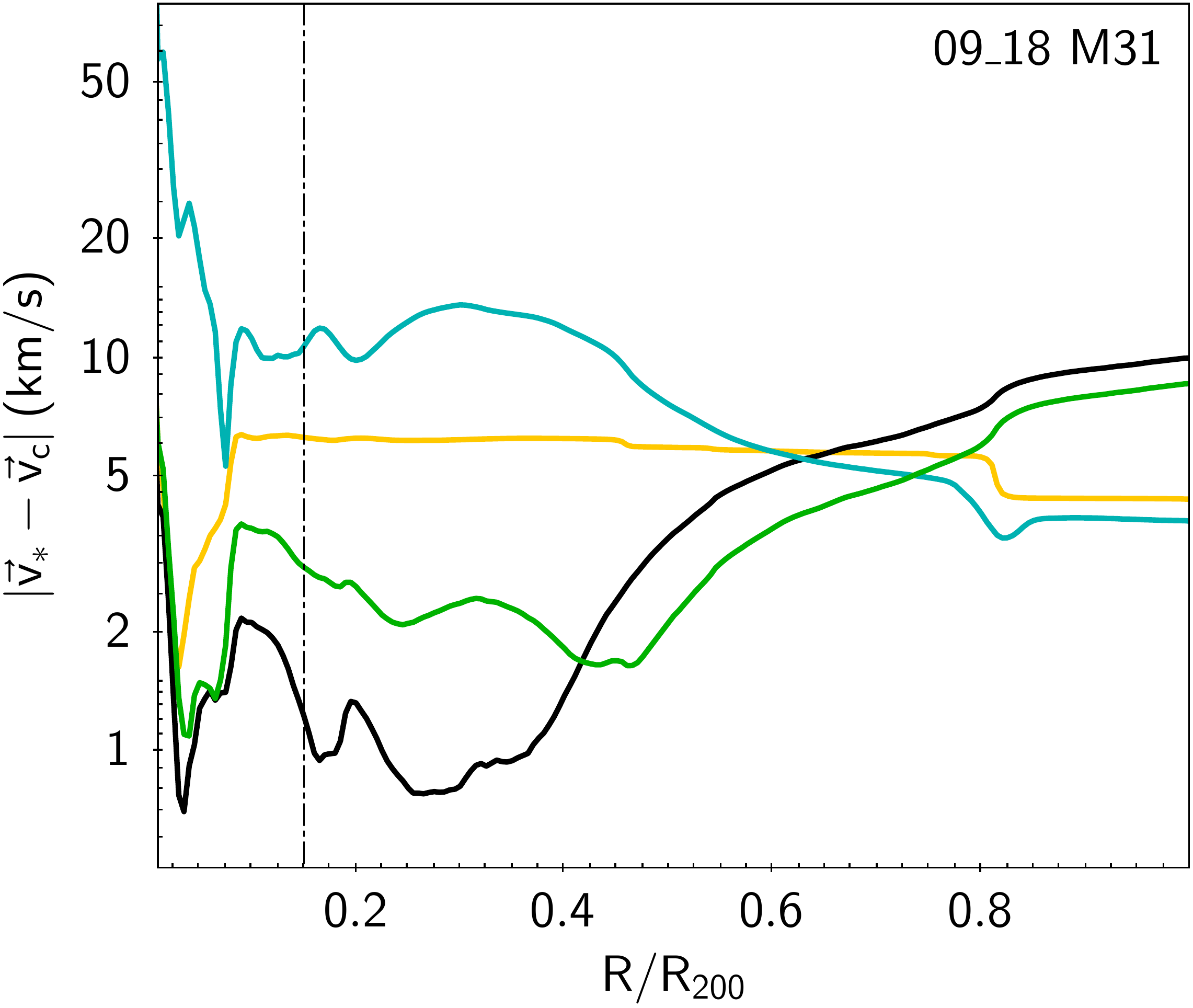} &
    \includegraphics[width=5.6cm]{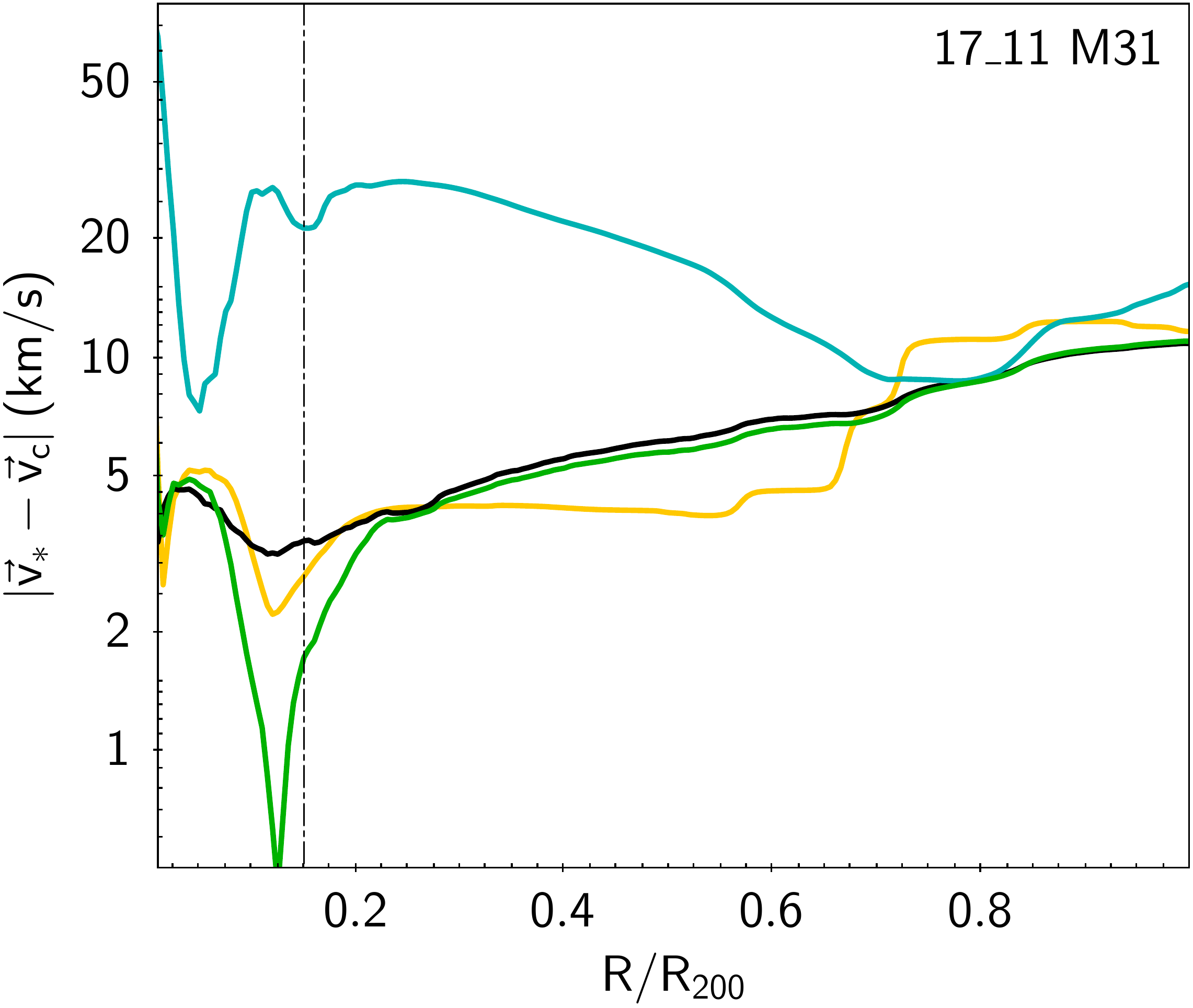} & \includegraphics[width=5.6cm]{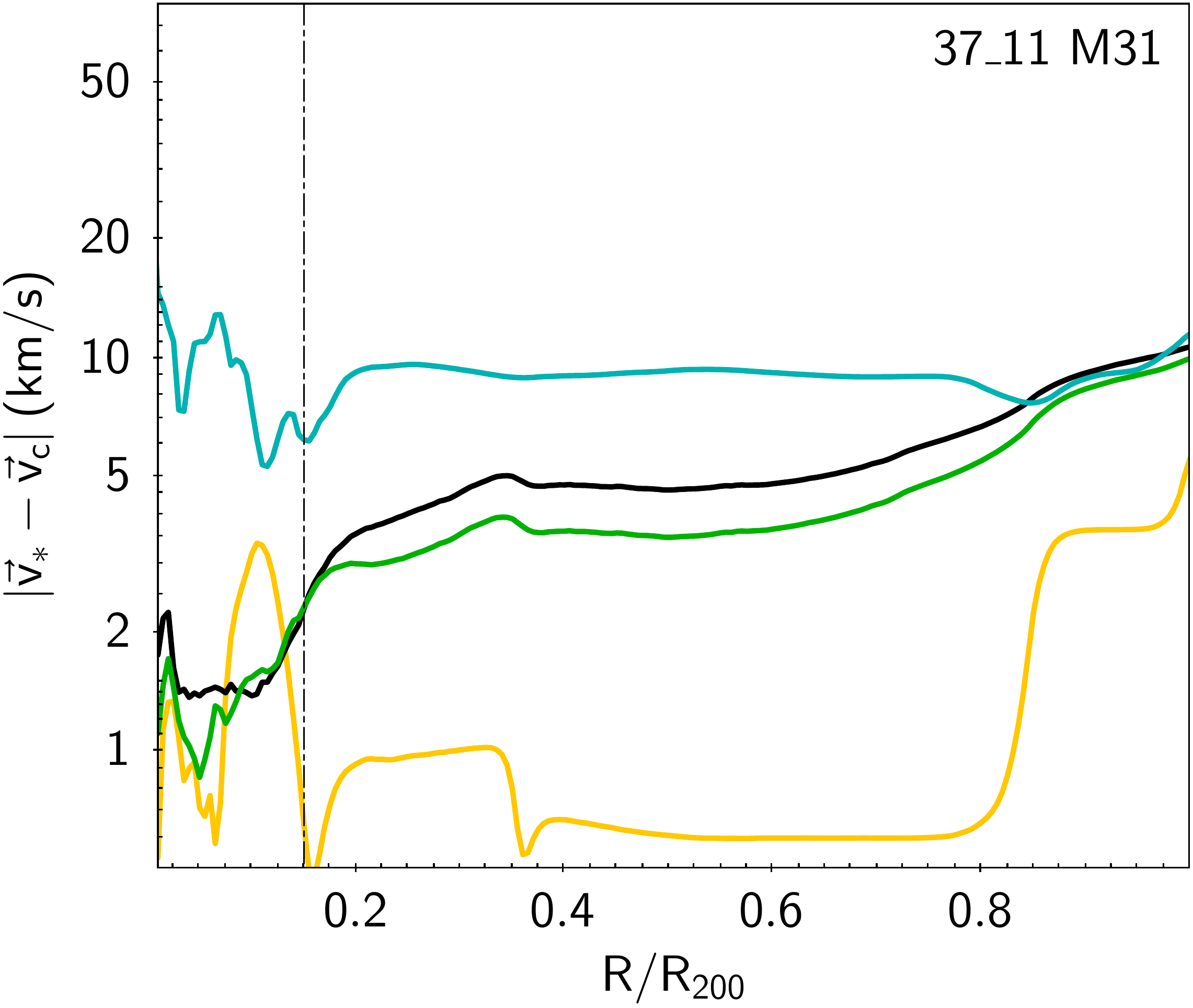} \\
    \includegraphics[width=5.6cm]{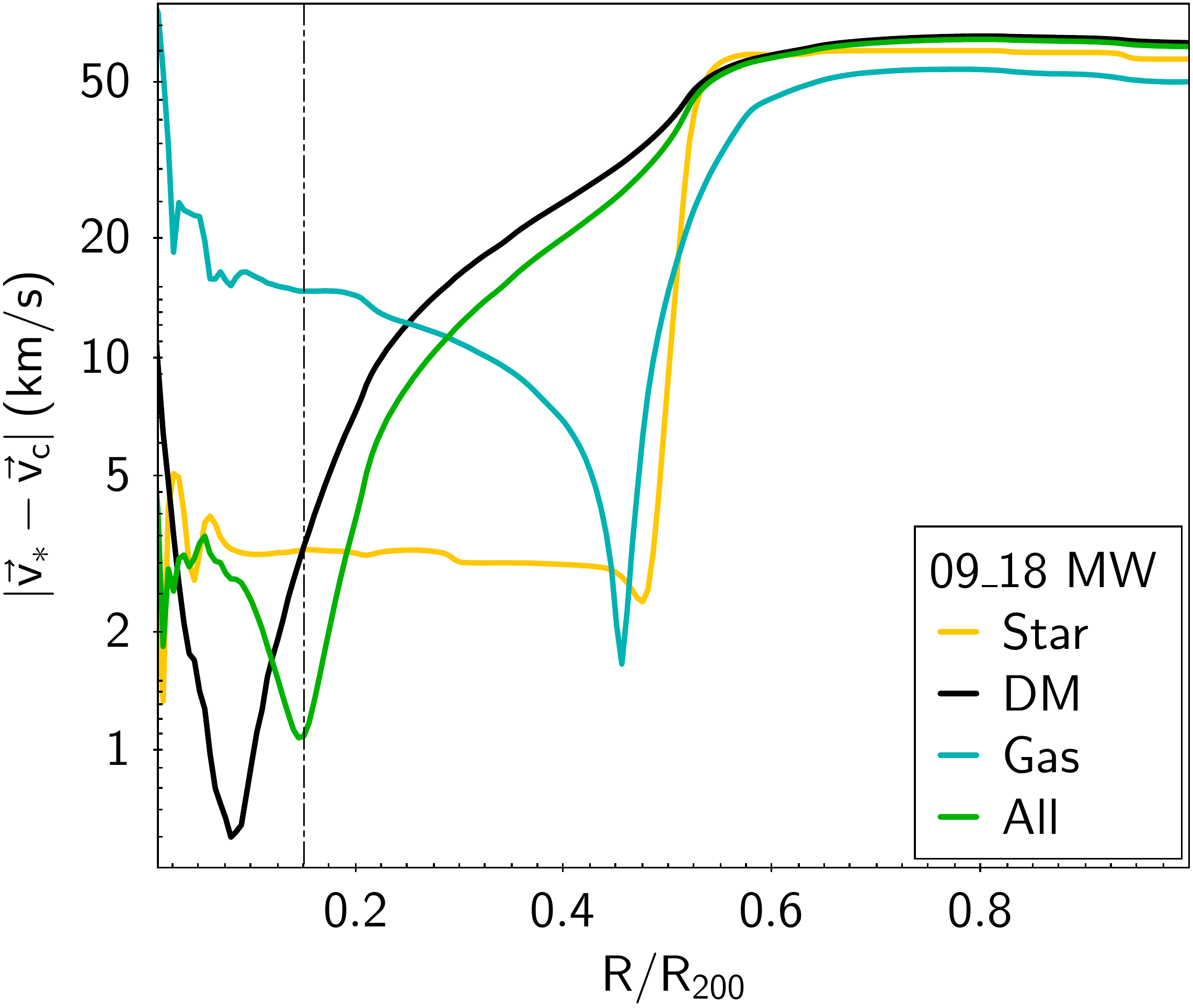} &
    \includegraphics[width=5.6cm]{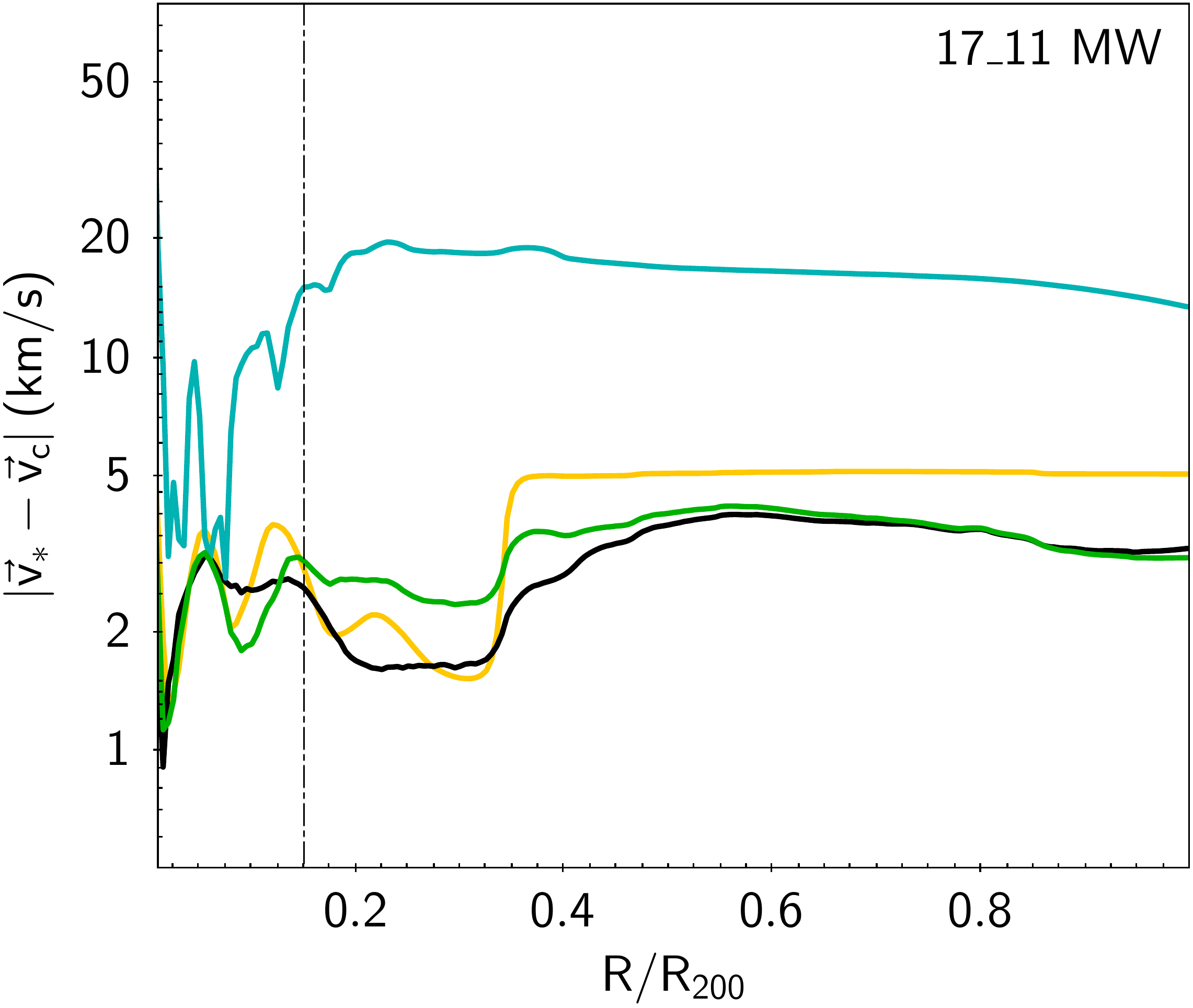} & 
    \includegraphics[width=5.6cm]{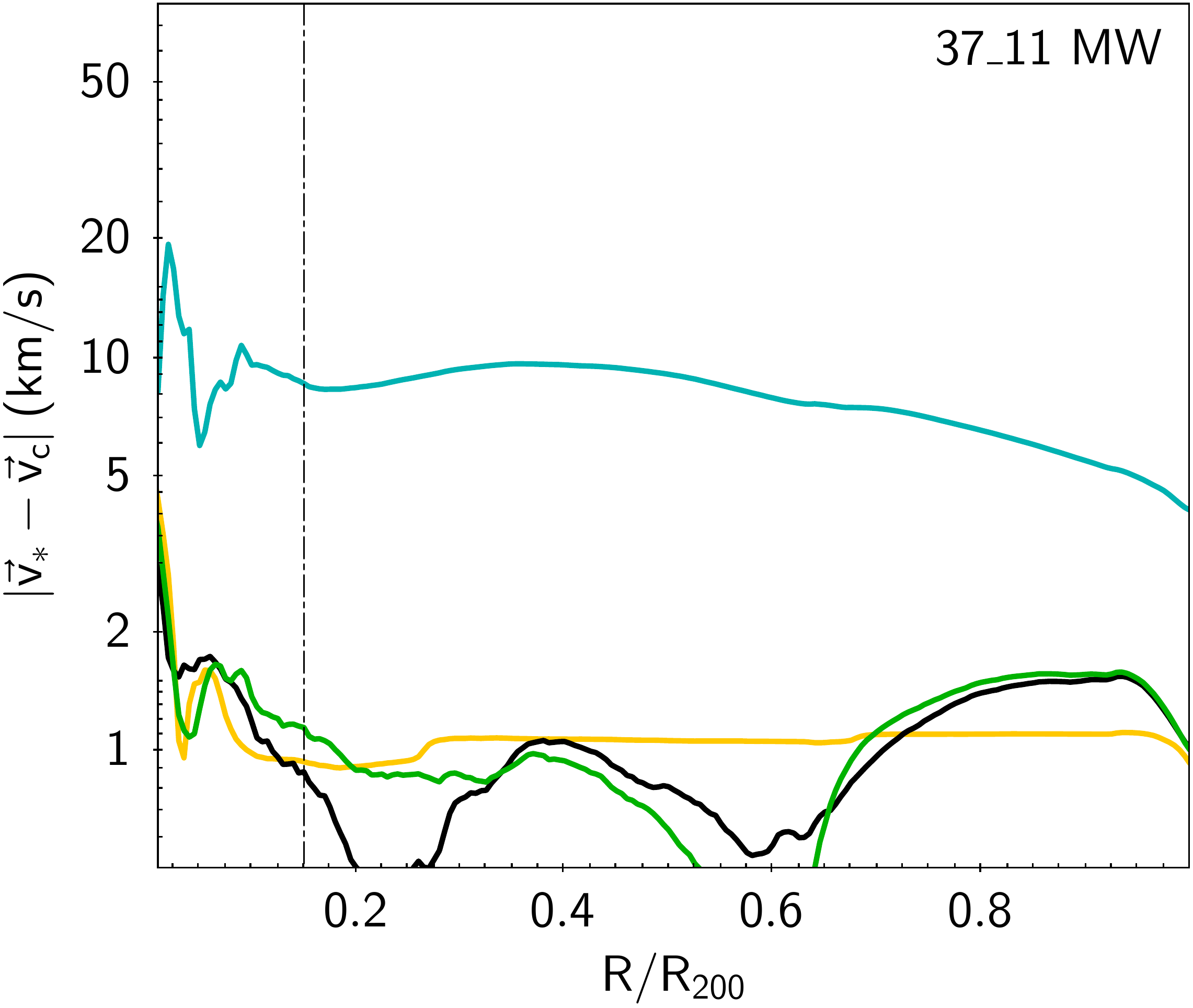} \\
   \end{tabular}
\caption{Same as Figure~\ref{diffx_part} but for the absolute differences in velocity between the disc centre and the centres of mass of different types of particles (equivalent cells). Note that given the uncertainties, any offset of less than 6 $\kms$ should be treated with caution (see Section \ref{discCentre}).}
    \label{diffv_part}
\end{figure*}

The velocity shifts (Figure~\ref{diffv_part}), on the other hand, can give rise to slightly different conclusions.
The velocity gaps between the edge of the disc and its centre are kept between 5 and 10$\kms$ for the stellar particles, at the limit of our confidence threshold in the detection of velocity gaps (established at 6$\kms$ in section~\ref{discCentre}). These values are nevertheless not surprising and remain within the range of expected and observed variations of the velocity dispersions for a spiral galaxy (e.g. \citealt{Robin22}). Over the same radius range, the velocity offsets for dark matter follow similar fluctuation amplitudes. 
This allows us to observe that baryons are tracers of the overall dynamics of the disc as they overwhelm the mass budget in the central part. When particles outside the disc and up to R$_{200}$ are considered, the velocity fluctuations are usually not larger. The shifts are in fact of the same order of magnitude as in the disc with sometimes a very slight increase. This is also true for the two simulations of 17\_11 and 37\_11 (M31), which, despite having a maximum positional offset of more than 10 kpc, do not show excessive velocity shifts.

%

There is one noticeable exception, the 09\_18 simulation for the MW, where the COM velocity of each type of particle, and thus of course also of all particles, diverges drastically from the COM velocity of the disc. 
Indeed, in this simulation, the central galaxy experiences a non-negligible accretion event. A sinking satellite comes within about 130 kpc of the host, clearly visible on the top right of the disc in Figure~\ref{img_simu_m31}. The total mass ratio between the host galaxy and the satellite is about 1:10. It is massive enough to make the velocity of the COM of the ensemble of particles deviate by more than 60$\kms$ at R$_{200}$ with respect to the disc velocity. The positive point is that this deviation is also traced by the stellar particles and could therefore be observed, analysed and corrected if necessary. Although the satellite in this simulation does not have exactly the same characteristics as the Magellanic clouds, it is nevertheless interesting to mention that this velocity offset between the disc of the 09\_18 (MW) simulation and the COM of its halo of 66$\kms$ is almost equal to the recent measurement of the velocity of the MW disc relative to its satellites, and about twice the preferred velocity of the disc relative to its halo \citep{Petersen21}. According to them, this disc travel velocity would be mostly caused by the recent passage of the Large Magellanic Cloud which is approaching the Galaxy and is about at 50 kpc from it.

In this first part, the COM behaviour of each type of particle has been analysed as a function of the distance to the centre of the disc. From the global point of view, it is reasonable to consider as true the hypothesis that star particles trace the behaviour of dark matter, and thus of the system as a whole, both in terms of position and kinematics of the barycentre. Of course, at a finer level, stars are more subject to rapid fluctuations compared with dark matter which is inherently more smoothly distributed. 

Given the very nature of gas cells in the simulation, they follow the behaviour of the other particles only very marginally with large variations. 
This has nevertheless a very limited impact for the COM calculation which remains steady in spite of the stirring of the gas. Indeed, at redshift zero, gas contributes in average to about 7\% of the total mass enclosed in the R$_{200}$ radius, about the same as for the amount of stars ($\sim$6\%). But the gas distribution is more scattered and simultaneously smoothly distributed, being at any radius widely dominated by another component. Within the central part of the disc where the gas fluctuations are most pronounced (inner 20 kpc), the gas contribute to less than 20\% to the baryonic mass budget. 
When considering a broader volume, the mass of the gas is at most half the stellar mass at the disc edge. At even larger radius, when the halo is lacking stars, the gas is very diffuse and its mass is staying bellow 10\% of the dark matter mass. The presence of satellites do not change drastically the amount of gas since they are mostly gas deficient.

Up to the galactic radius, the discrepancies in position between the different COMs and the disc, remain moderate.
But beyond, contrary to what one might expect and in most of the simulations, the larger the number of particles considered, i.e. the larger the radius, the larger the divergences between the COMs and the disc. This is in line with the results of \cite{GaravitoCamargo21}.
With the help of N-body simulations of the MW and the LMC, they find no offset in the inner halo (<30kpc) but note differences in position of up to 15 kpc at larger radii. This means that the centre of the disc and that of the whole halo do not correspond, the two points are indeed not equivalent. There is thus an actual decoupling between the central disc and the host halo.  
The positional offsets become substantial at R$_{200}$ for half of our sample. The accretion history and the size of the mergers impact a lot the instantaneous position at redshift 0, which can be shifted towards the outskirts of the disc or even outside of it.

However, the kinematics exhibit a slightly different behaviour.
At the particle level, there are no significant differences between the velocities of the COMs and the disc velocity which actually do not exceed the fluctuations of the velocity dispersions intrinsic to a spiral galaxy. But this is only true when there is no sizeable merger. Indeed, the observed differences can be important in the case of a prominent accretion, and in fact of the same order of magnitude as the observational shifts observed between the MW disc and its halo \citep{Petersen21}, deviations attributed to the effect of the Magellanic clouds.
It is also interesting to note that even if our goal is not to recover or to compare the exact configuration of the MW and the LMC, the deviations found in the 09\_18 (MW) simulation are not far from those obtained using N-body simulations modelling this dipole. 
While we find the values (30 kpc, 66$\kms$), respectively for the deviations in position and velocity, \cite{Gomez15} find the values (30 kpc, 75$\kms$) and \cite{Petersen20} a deviation in velocity of 40$\kms$.
We can then conclude that in the case of a major accretion, the N-body simulations are in agreement with our cosmological simulations. This means that the impact of a massive satellite is such that it gravitationally dominates, at least at the time of its accretion, all other events, in particular the galaxy formation history and the more minor accretions. Therefore, the study of this kind of interaction is sufficiently well reproduced by the N-body simulations without the need to use cosmological simulations.

\section{Centres of mass of subhaloes and satellites}

\subsection{Subhaloes}\label{sect:COMshalo}

The aim here is to examine the behaviour in the phase space of the barycentre of the host galaxy's subhaloes. Only those previously identified by AHF as gravitationally bound to the main halo and within R$_{200}$ are considered. Depending on the mass of these subhaloes, populations are defined and studied as a single object, in terms of its COM characteristics. The objective is to evaluate and quantify the position and velocity shifts that may exist between the central disc and its subhalo cohort. In addition to shedding light on the equilibrium state of the system as a whole, the amplitude of these deviations will indicate whether the subhaloes are good tracers of the kinematics of their host galaxy.

\begin{figure}
	\includegraphics[trim = {0cm 2.1cm 0cm 0cm}, clip, width=\columnwidth]{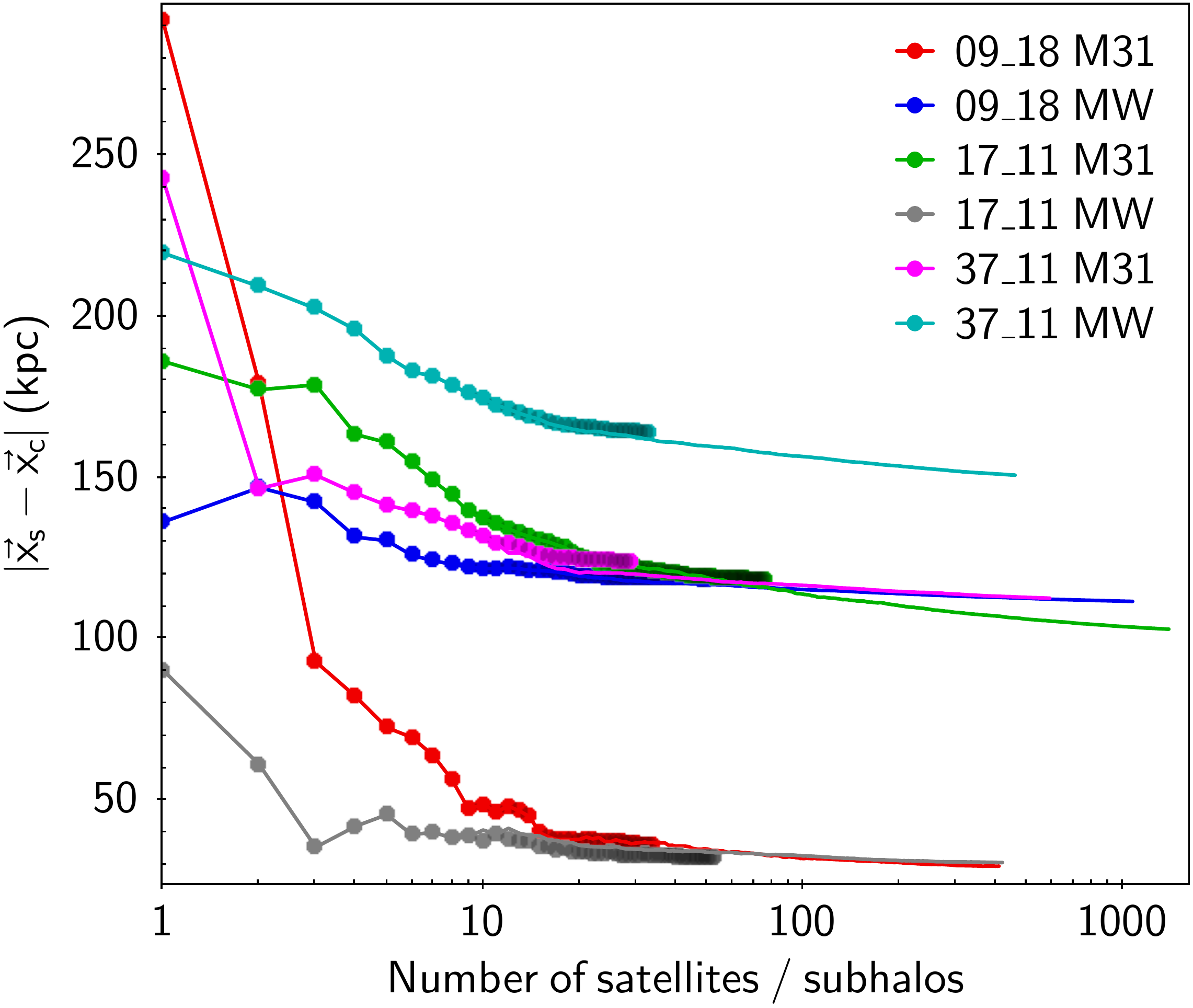}
	\includegraphics[width=\columnwidth]{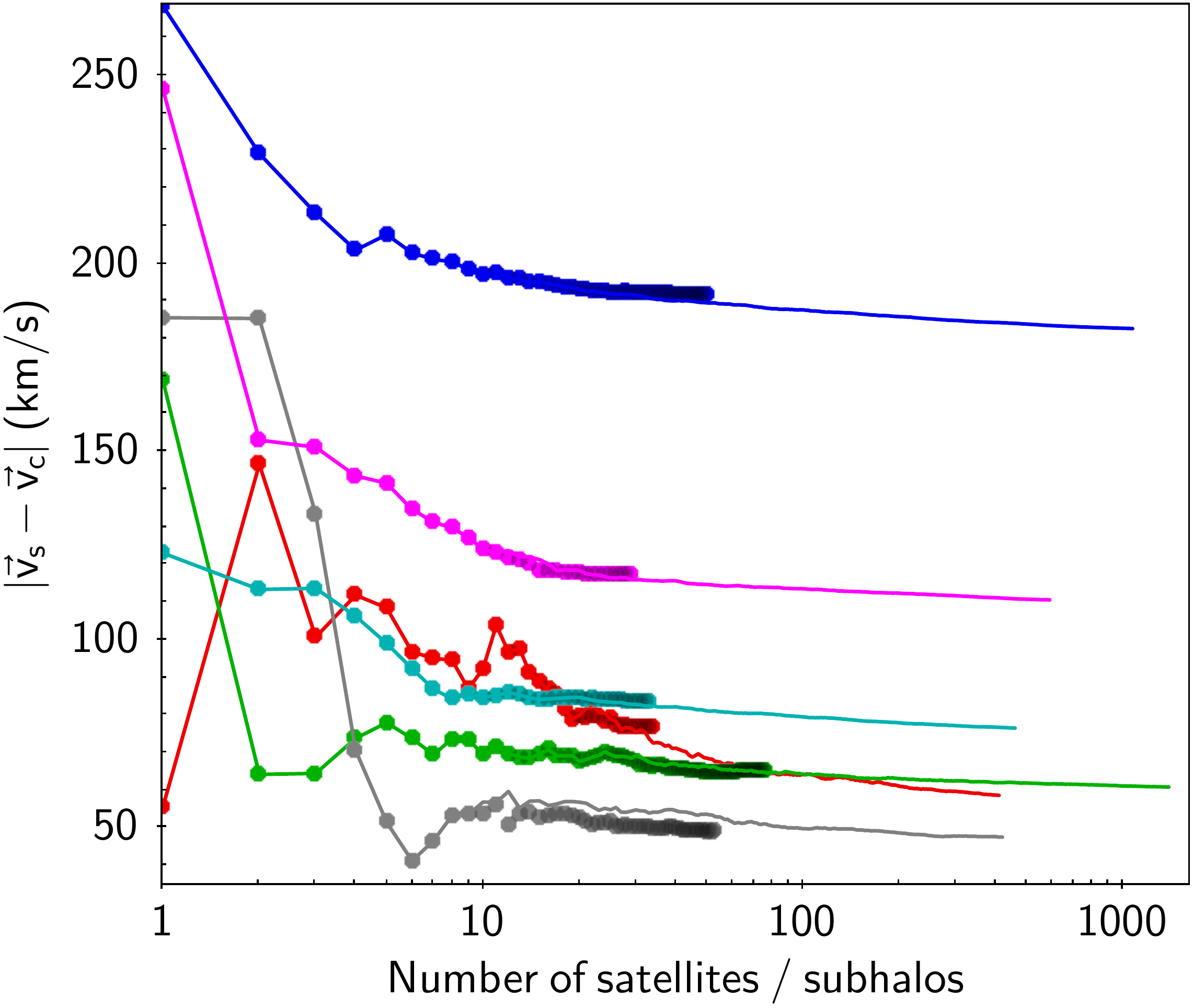}
    \caption{Absolute differences in position (upper panel) and in velocity (lower panel) between the disc centre and the centres of mass of various populations of satellites only (dots) or all subhaloes (plain lines). Those differences are given with respect to the number of most massive satellites or subhaloes taken into account. Results are colour coded according to the six host galaxies.
    }
    \label{diffxv_sathalo}
\end{figure}

\begin{figure}
	\includegraphics[width=\columnwidth]{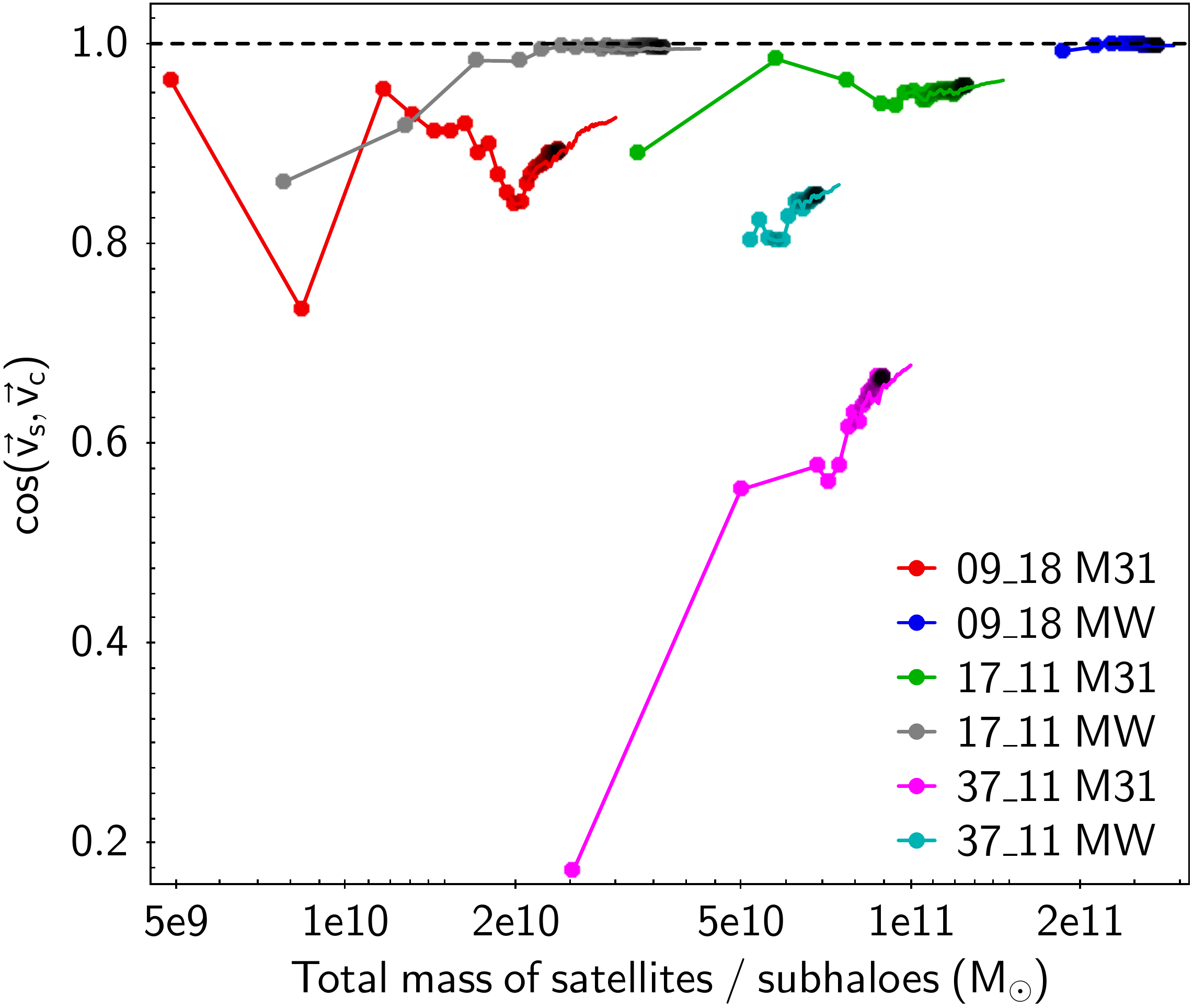}
    \caption{Deviations in orientation between the velocity vector of the disc centre and the velocity vector of the centres of mass of groups of satellites (dots) or subhaloes (plain line). The same populations as in Figure~\ref{diffxv_sathalo} are considered. Those differences are given with respect to the total mass of the group of satellites or subhaloes taken into account. The horizontal dashed line is a visual marker to point out a complete alignment.
    }
    \label{diffteta_sathalo}
\end{figure}

For each of the six hosts, the total number of subhaloes is considered and the COM calculated. Then, smaller and smaller subhalo populations are constructed by successively removing the least massive subhalo. The position and velocity of the COM are each time compared to our reference - the COM characteristics of the central disc. The solid lines in the Figure~\ref{diffxv_sathalo} summarises the positional shifts (top) and velocity differences (bottom), with respect to the number of subhaloes considered.
We also give in Figure~\ref{diffteta_sathalo} the angle between the velocity orientation of the disc and that of the COM of the subhaloes in plain line with respect to the mass of the population considered.
As an example, a population of 5 subhaloes (x axis in Figure~\ref{diffxv_sathalo}) means that only the 5 most massive subhaloes of the host are considered when computing ($\vec{\rm x}_{\rm s}$, $\vec{\rm v}_{\rm s}$). It corresponds to the fifth point from the left in Figure~\ref{diffteta_sathalo}. This figure also allows us to indicate the total mass of the subhalo populations, by looking at the x-coordinate of the right-hand end of each line.

%

%

It can first be seen that for all the simulations and for each of the parameters studied, the evolution of the shifts is similar as the number of subhaloes considered increases. After a phase where the variations are erratic when the few most massive subhaloes are used for the COM calculation, the general trends all reach a plateau fairly quickly.
This asymptote, slightly decreasing for the deviations in position and velocity, and slightly increasing for the cosine of the angle between the orientation of the velocity vectors, is already attained after taking into account only about a dozen of subhaloes. Moreover, this behaviour does not depend on the mass of the host galaxy, nor on the total mass of the subhalo populations nor even on the mass of the most massive subhaloes.

On the other hand, it is obvious that the COMs of the different subhalo populations selected do not follow the COM of the host galaxy disc. The deviations are large, heterogeneous, and non-negligible, considering for example the size of the disc as well as its rotation velocity. The differences in positions at the plateau cover an interval between about 30 and 150 kpc, those in velocities between 33 and 182$\kms$ and in velocity vector orientation between 0 and 47 degrees. Such deviations are an indication that the different subhalo populations, whatever the mass detection limits considered, are not in a dynamical state of relaxation with respect to their central galaxy. The magnitude of the offsets and their scattering depending on the host studied reflect the non-isotropic nature of the satellite populations. The differences are sufficiently large to explain the scattering of the results in the calculation of the MW, M31 and LG masses from the application of the virial theorem \citep{Diaz14, Hartl22}.

Looking at each galaxy separately, we can see that the three most massive host haloes, namely 17\_11 (MW), 09\_18 (M31) and 17\_11 (M31) are the least prone to being out of phase in both position, velocity and orientation. This could mean either that these galaxies have had time to fully accrete more sub-structures, or that they hold a more homogeneous satellite population. In particular, subhaloes are at present time either blended with the central galaxy - the 17\_11 (MW) and 09\_18 (M31) hosts have the smallest number of identified subhaloes and the lowest sub-population mass - or numerous and relaxed enough to have a fairly uniform phase space distribution - the 17\_11 (M31) host has the largest number of identified subhaloes.
This interpretation is supported by the shape of the stellar halo around the central disc (see Figure~\ref{img_simu_m31}), which is much extended for these three galaxies, attesting to numerous accretions. 

The most decoupled galaxies in terms of position and velocity, respectively 37\_11 (MW) and 09\_18 (MW), both have a sub-population gravitationally dominated by one single satellite. It thus governs the properties of the COM, notably the difference in orientation of the velocity vector. The blue and cyan curves in Figure~\ref{diffteta_sathalo} are very tight. Nevertheless, their configuration is not as identical. In the 37\_11 (MW) simulation, the massive satellite is far away, at the limit of the R$_{200}$ radius. It has little impact on the central disc. This is not the case for the 09\_18 (MW) simulation, where the massive satellite is closer and has already passed nearby the centre. It has then impacted the central disc, imparting to it a velocity aligned with its accretion direction. As a result, the orientation of the COM velocity of the subhaloes is perfectly aligned with that of the disc.

\subsection{Satellites}\label{sect:COMsat}

It is now necessary to examine whether the deviations found for the subhaloes, intrinsic to the dynamics of the complexes comprising a host galaxy and its substructures up to R$_{200}$, are still visible and of the same amplitude when only satellites are considered.
In other words, do the shifts on the satellites - which can ultimately be observed - depict correctly those of all the subhaloes? 

Thus, the second main population studied is the satellites. From the whole set of subhaloes gravitationally bound to their respective host previously selected, only the subhaloes containing at least one stellar particle are kept. And the similar exercise to the previous one is performed. For each host galaxy, the COM of the satellite population is calculated and compared to that of the central disc. Then the least massive one is removed and the sequence is repeated until only one is retained. The satellite populations considered now are of course sub-populations of the previous samples.

The results of the different deviations are reported in Figures~\ref{diffxv_sathalo} and~\ref{diffteta_sathalo} as dots.
The vast majority of the most massive subhaloes are systematically populated by stars and identified as satellites. As a natural consequence, the general profiles of deviations remain almost similar to those of the subhaloes. The same implications can be drawn from this.
Only slight exceptions can be visible as for example with the simulation 17\_11 (MW) where the twelfth most massive subhalo is dark (see the difference between the twelfth gray dot from the left and the line in Figure~\ref{diffxv_sathalo}). But even in those cases, the trends continue to be identical. In the range of mass of the smallest satellites, some subhaloes can be, or not, lightened. That is the reason why the lines in the figures can be slightly misaligned with the dots when the curves become flatter. Nevertheless, the plateau identified with the subhaloes is still similarly reached with satellites, of course with less objects. Thus, the position and the kinematics of the COM of the few tens of satellites are almost identical to those of several hundreds of subhaloes.
The differences in position and velocity orientation between all subhaloes and satellites only can be of the order of 10 kpc and 5 degrees respectively, but never greater and always sporadic. The same is true for the velocity offsets, which are typically under 10$\kms$. 
There is only one slightly more pronounced offset, of 20 $\kms$ for 09\_18 (M31), which is one of the most massive and least perturbed hosts at z=0. The total mass of its substructures is the smallest among our six simulations. Furthermore, the mass of its subhaloes is 25\% more than the mass of its satellites, which explains the relatively stronger influence of the dark haloes. It can therefore be noticed that the latter are cooler as they allow to lower the differences between between the population of subhaloes and the disc.

It follows that the study of the populations of more than about a dozen of the most massive satellites, dominating the environment of the host galaxy, is a sufficient indicator to study the kinematics of the entire population of subhaloes. The COM properties of the satellites provides a upper limit to those of the COM of the whole substructures within the host halo. It is therefore reasonable to consider that the observational data, even with their biases and completeness problems, are sufficient to allow us to understand precisely the behaviour of the COM of the MW and M31 galaxy substructures.

\section{Influence of the most massive satellites}\label{sect:COMmassive}

\begin{figure*}
    \centering
        \begin{tabular}{cc}
    \includegraphics[trim = {0cm 2.45cm 0cm 0.0cm}, clip,width=8.65cm]{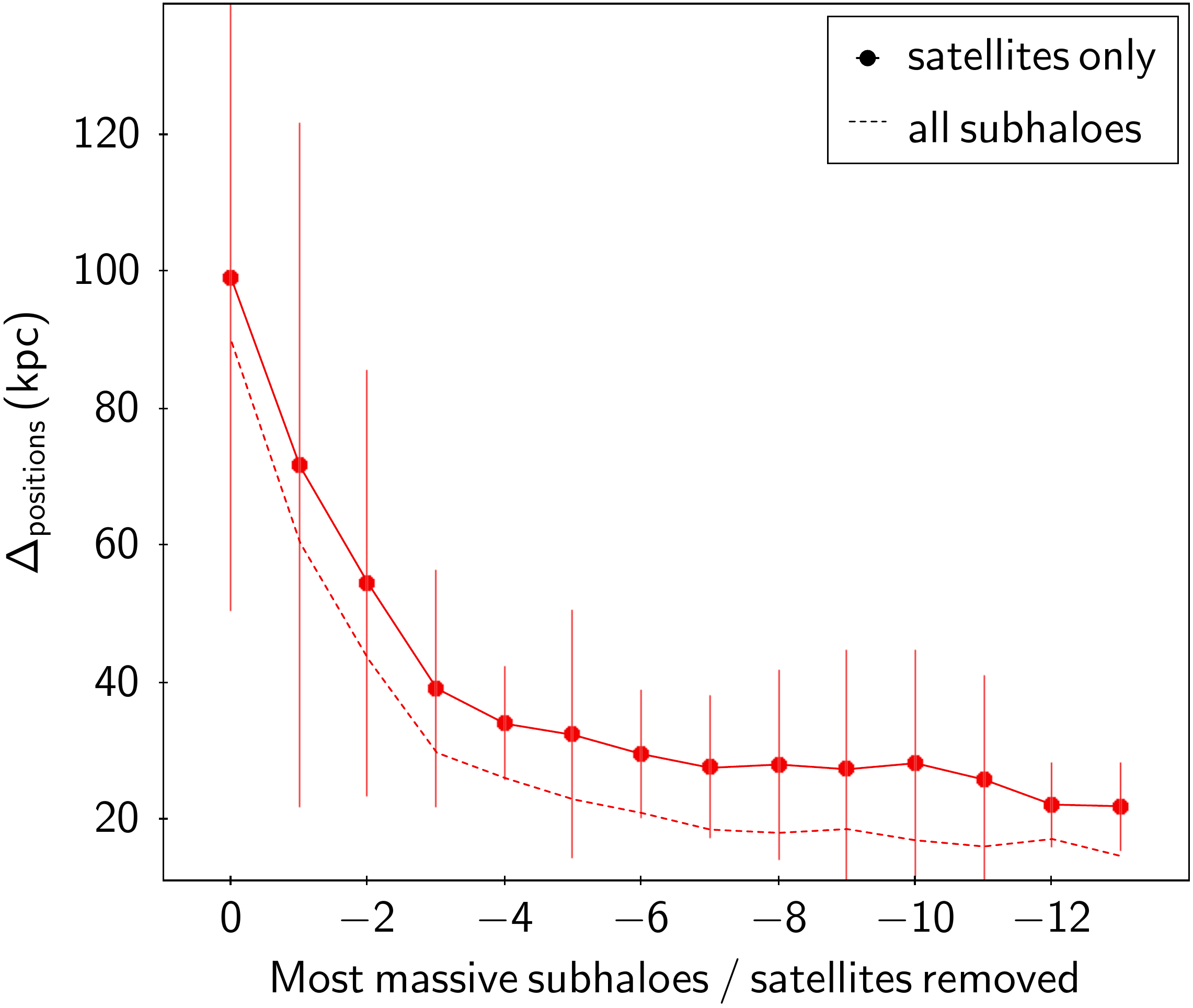} & 
    \includegraphics[trim = {0cm 2.45cm 0cm 0cm}, clip,width=8.65cm]{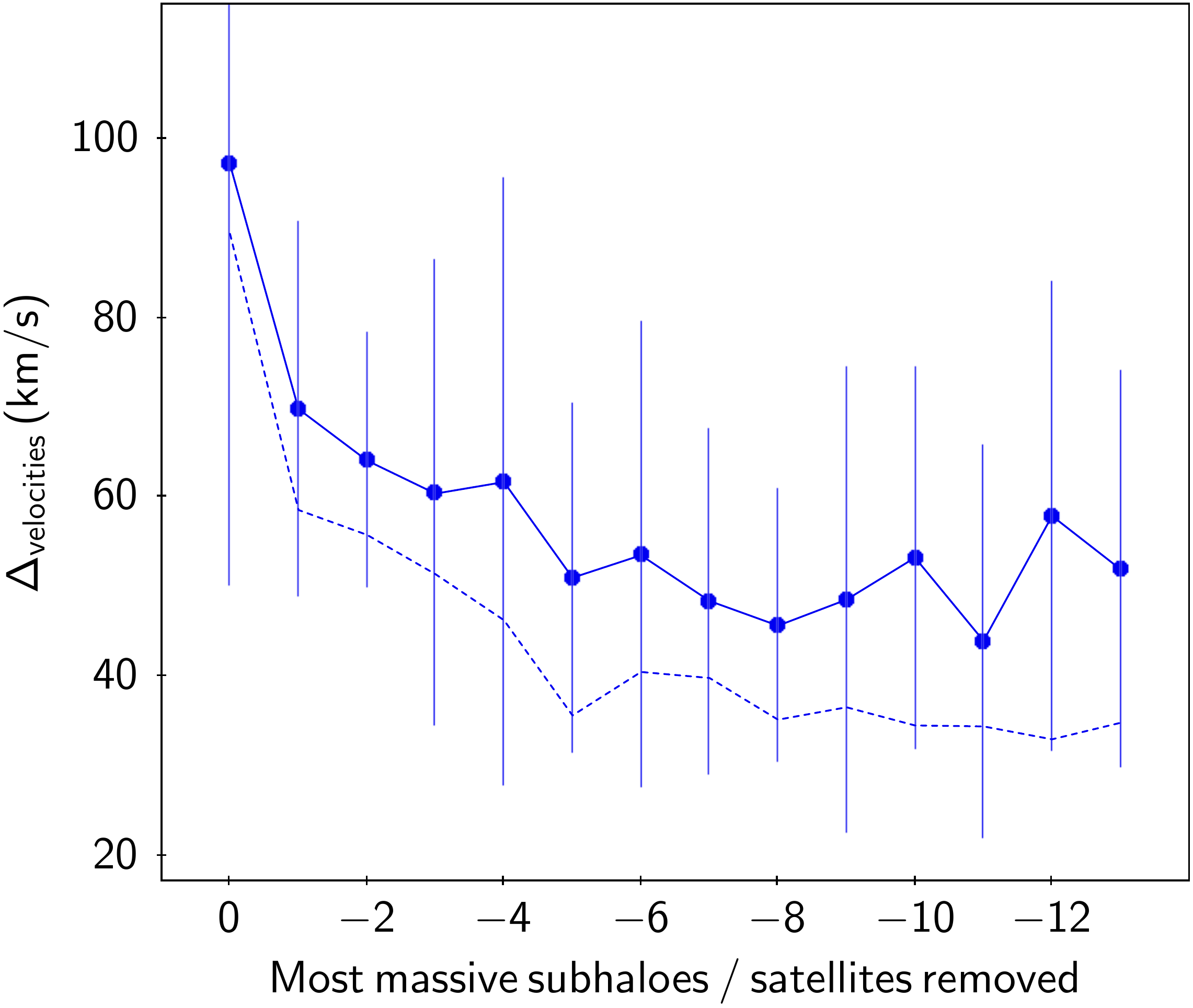} \\
    \includegraphics[trim = {-0.45cm 2.15cm 0cm 0.05cm}, clip,width=8.65cm]{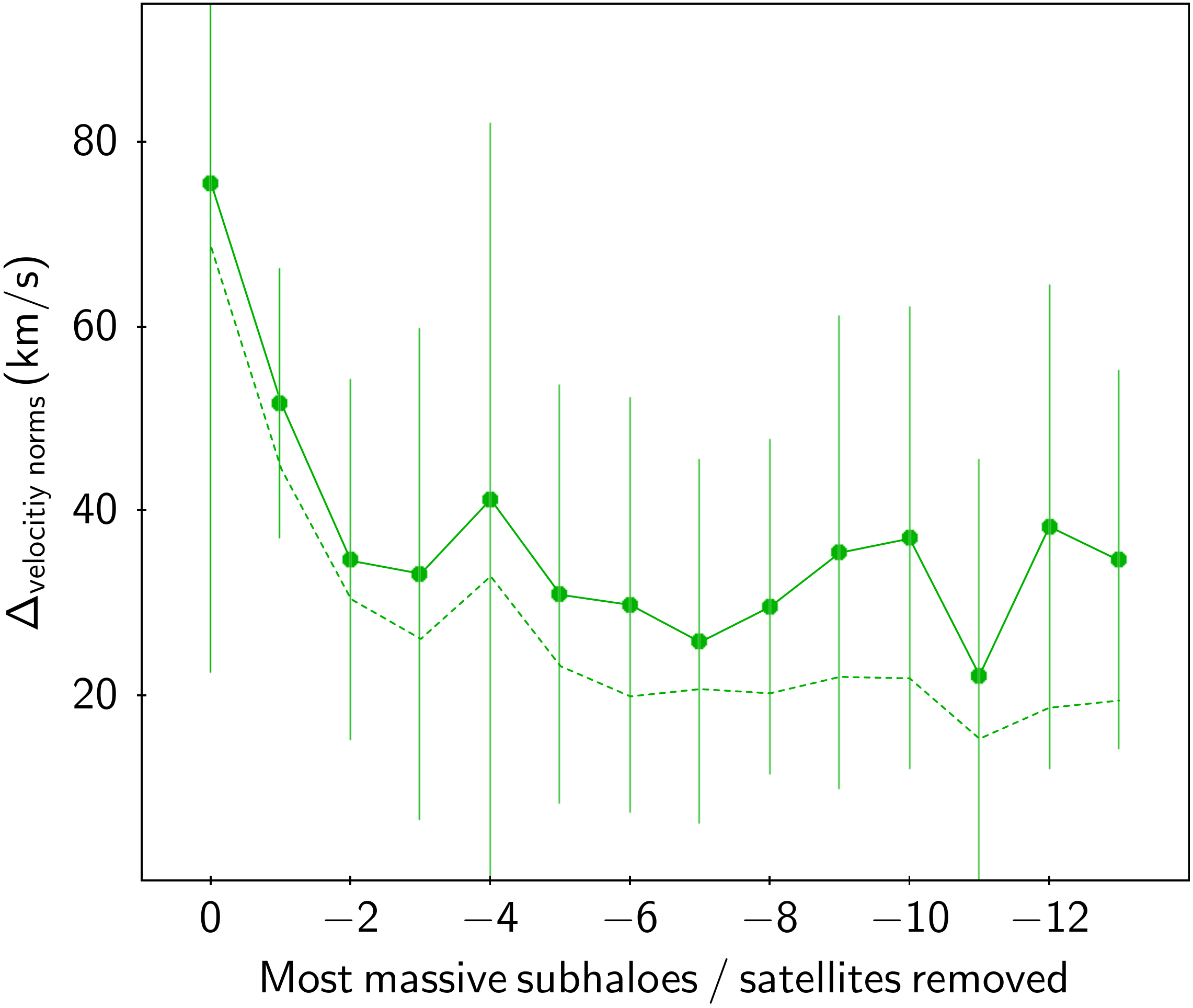} & 
    \includegraphics[trim = {-0.25cm 2.15cm 0cm -0.4cm}, clip,width=8.65cm]{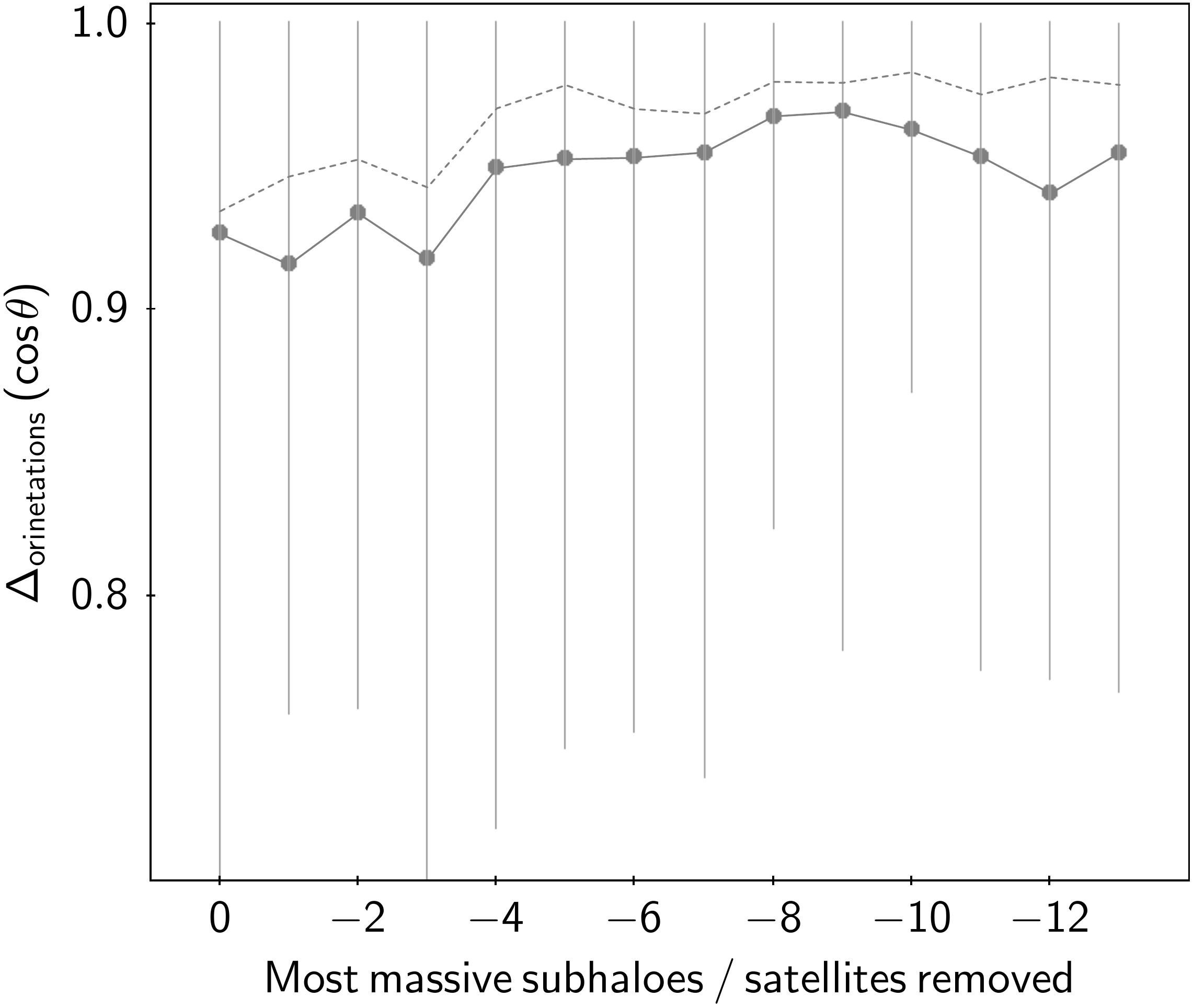} \\
    \includegraphics[trim = {-0.25cm 0cm 0cm 0cm}, clip,width=8.65cm]{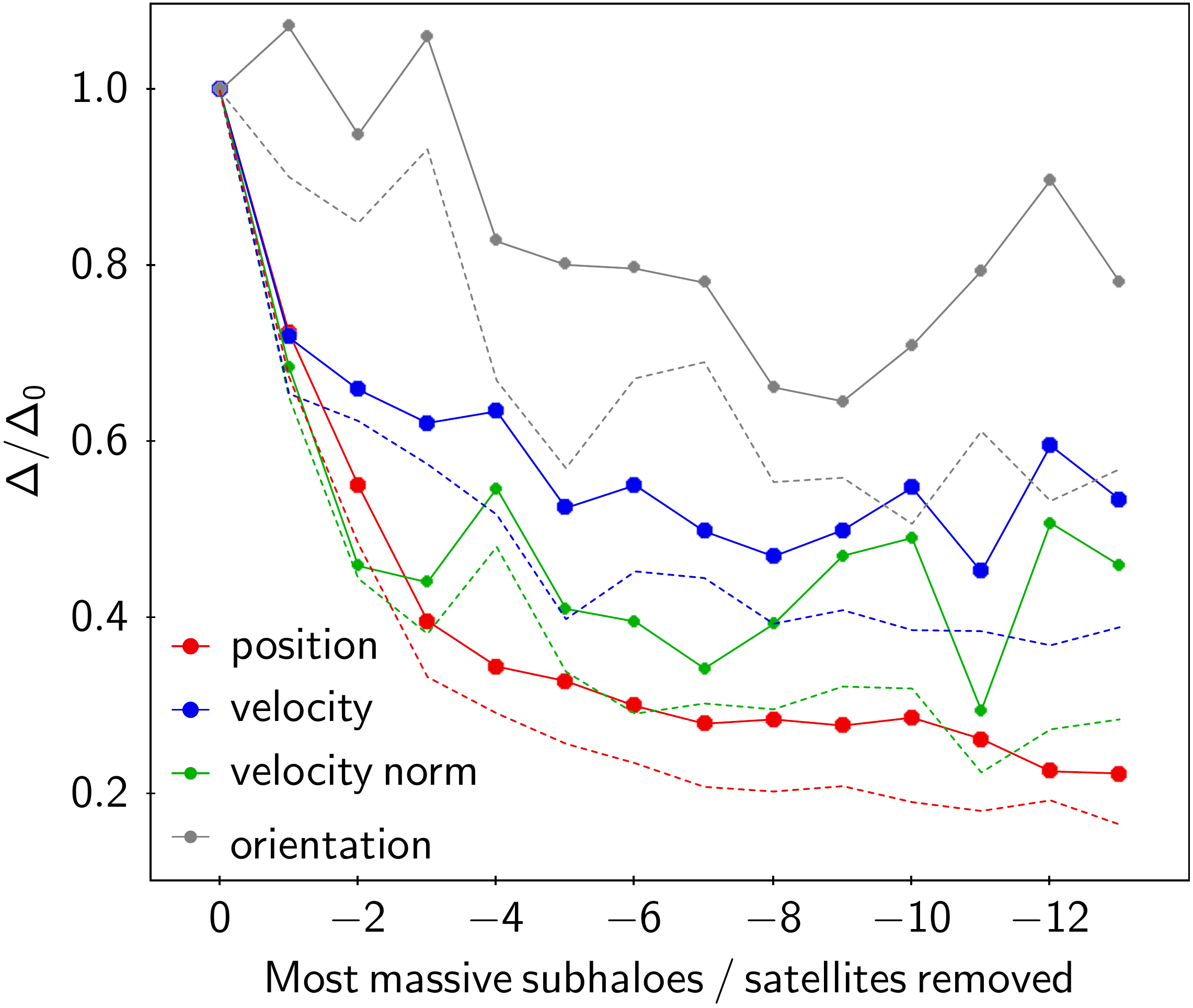} & 
    \includegraphics[trim = {-0.25cm 0cm 0cm 0cm}, clip,width=8.65cm]{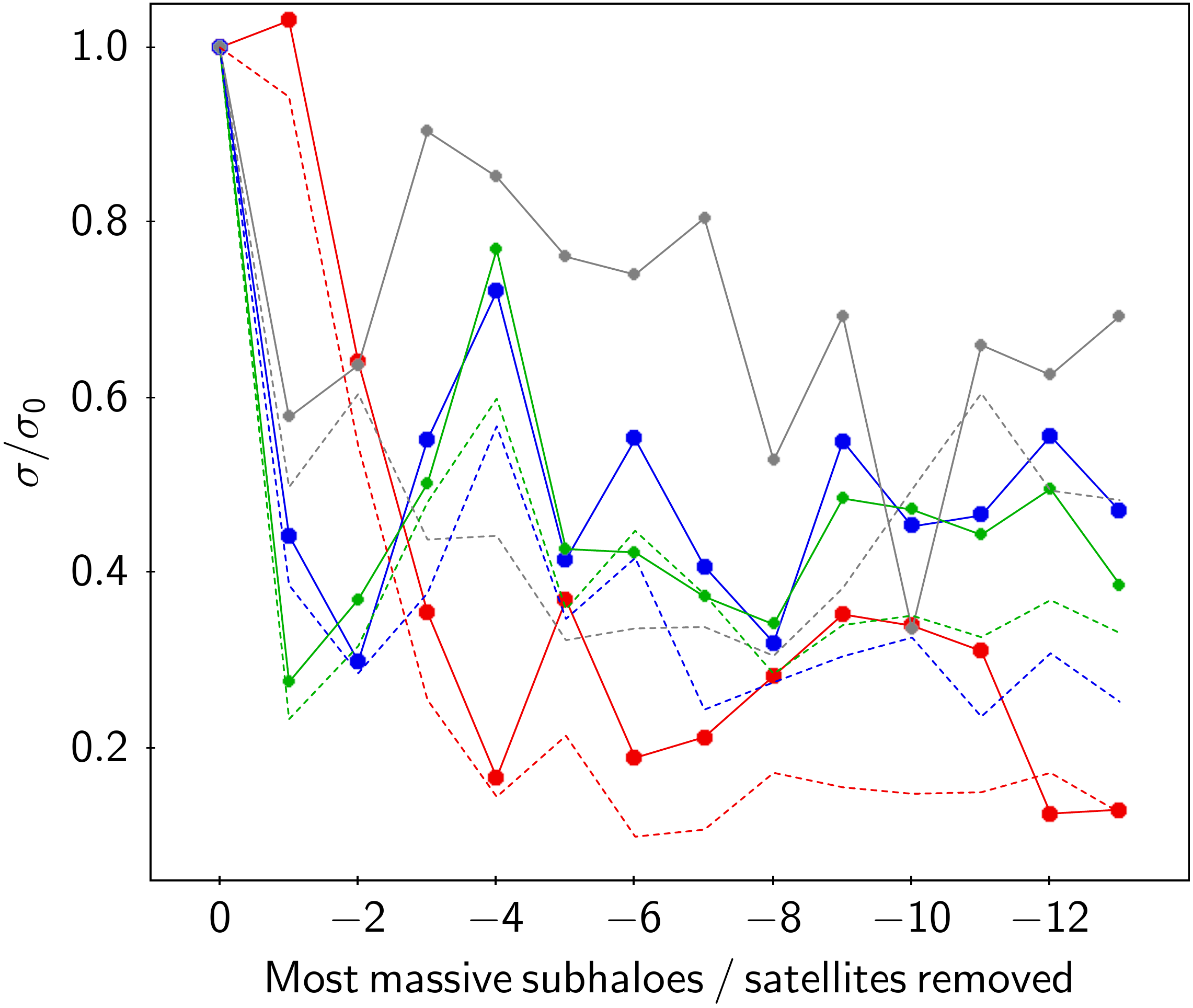} \\
   \end{tabular}
\caption{Averaged offsets for the six host galaxies between the COM of the central disc with that of the system of its satellites (lines and dots) or subhaloes (dashed line). The horizontal axis is graduated according to the number of the most massive satellites (or subhaloes) removed.
The four top panels present the mean evolution of the differences in positions (in red), velocities (in blue), velocity norms (in green) and velocity orientation (in grey). The uncertainties indicate the one sigma standard deviation for the satellite populations. The two lower panels summarise the relative evolution (normalised by the values when all satellites or subhaloes are considered) of the parameters investigated. The left-hand panel shows the evolution of the offsets and the right-hand panel shows the evolution of the uncertainties.
}
    \label{diffxv_sathalo_X}
\end{figure*}

The previous sections have revealed significant differences in position and velocity between the COM of the central galaxies and their respective satellite or subhalo populations.
We have also seen that a host perturbed by a massive satellite can exhibit stronger COM shifts, the most discernible example being the simulation 09\_18 (MW).
In a further perspective of transposing these results to observations, one could ask whether, by selecting a sub-population of satellites, it would be possible to contain these differences.
Thus, it is now important to investigate whether these deviations from the disc centre can be induced by a few satellites only, in particular the most massive ones.
In order to test this hypothesis, a scheme similar to the previous one is set up. For each of the six hosts, all the satellites are considered and the COM is calculated. Then, smaller and smaller populations of satellites are constructed by successively removing the most massive satellite. This test is also performed for the subhaloes. The differences between the COM of these populations and the disc, in term of positions and velocities, are derived.
The shifts are analysed statistically, taking into account the ensemble of the six hosts.
The four top panels of the Figure~\ref{diffxv_sathalo_X} show the average evolution of the deviations of the position, velocity, velocity norm and orientation as a function of the number of the most massive satellites or subhaloes removed. The bottom panels illustrate the relative decay of the offsets and their associated uncertainties.

%
%
%
Starting from the left side of the plots where all satellites (or subhaloes) are taken into account, the deviations all tend to decrease sharply before stabilising once the most massive objects are removed. 
For example, eliminating at least the five most massive satellites from the COM position calculation ensures that the maximum positional offset from the host will always be statistically below 30 kpc, which is only one third of the offset from the entire population. With the same restrictions, the velocity offset will always be under 60 $\kms$, representing only sixty percent of the initial offset. The standard deviations between the six simulations on the spread of the offsets are also considerably reduced, being at most a maximum of 38\% of the initial standard deviation for position and 55\% for velocity.
It can also be noticed that the differences in terms of orientation of the velocity vectors concede only a moderate improvement of 10\%. This is mainly due to the fact that the velocity of the disc and that of the COM of the full population of satellites are already rather well aligned. This alignment is even better with the subhaloes. But in addition, the confidence in this alignment increases when the bigger satellites are not taken into account since the uncertainties decrease further, by about 20\%.

The comparison between satellites and subhaloes is not surprising here, in line with the previous sections. Subhaloes and satellites curves present similar behaviour. Consequently, the previous results showing that the kinematics of the subhaloes is sufficiently well represented by that of the satellites remains true since all the differences with the subhaloes are included in the one sigma standard deviation of the differences with the satellites only. The non-equilibrium state of the halo is thus predominantly dominated by the few most massive satellites. The COMs of the lower mass population of satellites and subhaloes are much more consistent with that of the central disc than with the whole population.
This also means that the less massive satellites are not absolutely subjected to the more massive ones. And if some deviations exist for the lighter satellites, they are small enough to statistically compensate each other to some extend thanks to a better homogeneity of their distributions in the phase space.
By not considering the few most massive objects, the differences in position and velocity are reduced by a factor of about three and two respectively, as well as are the uncertainties. Nevertheless, the average offset in velocity of 60 $\kms$ after removing the most massive satellites remains relatively high in absolute terms. It has been shown in the case of the MW that a massive satellite can affect almost all the other satellites, regardless of their distance \citep{Battaglia22}. In addition, the less massive satellites, potentially accreted previously in small groups, will experience interactions between themselves. They will also evolve in a non-stationary potential and in an inhomogeneous halo \citep{DSouza22}. The satellites will thus deviate from an isotropic distribution as observed for the MW \citep{Makarov23}. The residual value of 60 $\kms$ that we obtain is very likely a reflection of the impact of the most massive satellites on the orbit of the other satellites coupled with the evolutionary history of each object.

\section{Summary and conclusion}\label{sect:COMccl}

We have studied cosmological simulations of three pairs of galaxies at redshift 0, in a configuration similar to that of the LG. They have been analysed as six independent hosts, but in a realistic MW or M31 environment. After defining the centre of mass of the disc of each galaxy, we compared the position and velocity of this reference to other centre of mass calculations. 

We first studied the behaviour of the systems at the particles level to understand the physical deviations that can arise between the disc and the different components which populate the halo. The evolution of the total COM follows that of the dark matter, which itself is well traced by the stellar component. It appears that the position of the halo's COM is shifted from the disc centre by more than ten kpc for half of our sample. These deviations are important regarding the typical size of the disc of a spiral galaxy and should be taken into consideration in any related study. In the absence of significant nearby merger at redshift zero, the velocity of the halo's COM remains close to that of the disc, with deviations of about 10$\kms$ which is of the same order of magnitude as the expected velocity dispersion for a MW-like galaxy. But under the effect of a major accretion, in our case a mass ratio of one tenth at a distance of $\sim$130 kpc, the velocity deviation goes up to 66$\kms$. It can be qualitatively compared to the recent minimum value of 32$\kms$ for the displacement of the MW disc with respect to its halo under the influence of the Magellanic Clouds \citep{Petersen21}.

In a second step, we studied the COM of the system of subhaloes and satellites in order to examine if they follow the dynamics of the host. No major difference is found between the subhaloes and the satellite populations. Deviations in the phase space between the host disc and the COM of subhaloes span a large range with average values of $\overline{\Delta}_{\rm position} = 98 \pm 47$ kpc and $\overline{\Delta}_{\rm velocity} = 96 \pm 46 \kms$ with the whole population of objects. These values remain qualitatively the same when only half of the most massive objects are taken into account. But the COM position and velocity of smaller samples, when more less massive objects are being ignored, drift further away from those of the host disc centre. 
On the contrary, when the few most massive satellites are ignored from the COM calculation, the latter tends to get closer to that of the disc with approximated average values of $\overline{\Delta}_{\rm position} \approx 30 \pm 15$ kpc and $\overline{\Delta}_{\rm velocity} \approx 50 \pm 20 \kms$.

The goal of this study was to know what are the expected 3D physical discrepancies in position and velocity of the centre of mass between the disc of a MW-like host galaxy and its halo. 
A deeper understanding of the existing offsets between the different centres of a galactic complex is becoming a necessity, in particular with the {\it Gaia} mission \citep{Gaia16}, which has opened up a very favourable period for the collection of data on the proper motions of the objects surrounding the disc. One of the most fundamental parameters, whose value is intrinsically linked to the dynamical tracers used and therefore to their COMs, is the mass of the galaxy (see e.g. \citealt{Monari18, Riley19, Slizewski22, CorreaMagnus22, Patel22}) 

Several confirmations and recommendations have resulted from this work. 
Firstly, at the particle scale as well as at the object scale, the stars faithfully trace the global dynamics of the studied systems which is itself intrinsically governed by dark matter. 
Secondly, significant physical misalignments in position and velocity between the host disc and the COM of the whole halo exist. Those decoupling are not systematic but can be important depending on the simulation considered, and hence on the equilibrium and evolution state of the system. The offsets tend to systematically increase as the volume studied around the host increases, from a sphere of radius equal to the size of the disc to a sphere of radius R$_{200}$.
Thirdly, when detailed information (at the particle level) would be missing as well as information on dark matter, it is nevertheless possible to have a rough idea of the degree of disruption by looking at the COM of the satellite population. For this, there is no real need to reach completeness in terms of number of subhaloes.
The COM derived from at least a dozen of the most massive satellites exhibits phase space properties that are close enough to those of the COM calculated with all satellites.
Finally, we have been able to see that the halo disturbance is mainly driven by a few of the most massive satellites. And these do not fully dominate the individual dynamics of the other less massive objects which mostly remain decorrelated.
As a consequence, the COM of the satellite population, after subtracting the three to five most massive objects, is relatively close to the position and kinematics of the host disc. 
The four points summarized in this last paragraph give us the keys for a better understanding in any future analysis of the dynamics of the MW (or M31)-like satellite system with respect to the central disc of their host galaxy.

\section*{Acknowledgements}

JBS would like to thank Oliver Newton for his help in handling the simulation data.

\section*{Data Availability}

Data products used and derived in this article will be shared on reasonable request to the corresponding author.



\bibliographystyle{mnras}
\bibliography{COM_Hestia} 

\bsp	
\label{lastpage}
\end{document}